\documentclass[12pt,reqno]{amsart}
\usepackage{graphicx,amssymb,amsmath}
\usepackage{amsaddr}
\usepackage[utf8]{inputenc}
\usepackage{epstopdf}
\usepackage{url,hyperref}	
\usepackage{bm}
\usepackage{color}
\usepackage{array}
\usepackage{enumerate}
\usepackage{doi}

\usepackage{theorem}
\usepackage[nodayofweek]{datetime}

\usepackage{natbib}
\setcitestyle{authoryear,open={(},close={)}}
\usepackage{hyperref}
\usepackage{etoolbox}
\usepackage{chemformula}
\usepackage{gensymb}

\renewcommand*{\b}{{\boldsymbol{b}}}

\renewcommand*{\v}{{\boldsymbol{v}}}

\renewcommand*{\Xi}{{\boldsymbol{\xi}}}

\newcommand*{\R}{{\mathbb{R}}}

\renewcommand*{\u}{{\boldsymbol{u}}}
\newcommand*{\uobs}{{\boldsymbol{u}}^{\rm{o}}}
\newcommand*{\Uobs}{{\boldsymbol{U}}^{\rm o}}
\newcommand*{\utruth}{{\bf{u}}} 
\newcommand*{\uvalid}{{\bf{u}}_{\rm{valid}}} 

\newcommand*{\Win}{{\boldsymbol{W}_{\rm{in}}}}
\newcommand*{\bin}{{\boldsymbol{b}_{\rm{in}}}}
\newcommand*{\Robs}{\bm{\mathrm{\Gamma}}}
\newcommand*{\W}{{\boldsymbol{W}}}
\newcommand*{\X}{{\boldsymbol{X}}}
\newcommand*{\Ntrain}{N}

\newcommand*{\Res}{{\boldsymbol{\Phi}}}
\newcommand*{\U}{{\boldsymbol{U}}}

\newcommand*{\w}{{\boldsymbol{w}}}
\newcommand*{\K}{{\boldsymbol{K}}}
\renewcommand*{\H}{{\boldsymbol{H}}}

\newcommand*{\z}{{\boldsymbol{z}}}
\newcommand*{\Pf}{{\boldsymbol{P}}^{\rm f}}

\newcommand*{\Z}{{\boldsymbol{Z}}}
\renewcommand*{\P}{{\boldsymbol{P}}}

\newcommand{\tauf}{\tau_f}
\newcommand{\CRPS}{\rm{CRPS}}

\usepackage[ruled,vlined,boxed,commentsnumbered]{algorithm2e}

\usepackage{moreverb}
\usepackage[colorinlistoftodos]{todonotes}

\newcounter{gagcomment}

\title{Supervised learning from noisy observations: Combining machine-learning techniques with data assimilation} 

\author[G. A. Gottwald]{Georg A. Gottwald}
\address[G. A. Gottwald]
{School of Mathematics and Statistics \\
 University of Sydney \\
 NSW 2006 \\
 Australia
}
\author[S. Reich]{Sebastian Reich}
\address[S. Reich]
{Institute of Mathematics\\
 University of Potsdam \\
Germany
}
\email[G. A. Gottwald]{georg.gottwald@sydney.edu.au} 
\email[S. Reich]{sebastian.reich@uni-potsdam.de}

\begin{document}

\maketitle


\begin{abstract}
Data-driven prediction and physics-agnostic machine-learning methods have attracted increased interest in recent years achieving forecast horizons going well beyond those to be expected for chaotic dynamical systems.  In a separate strand of research data-assimilation has been successfully used to optimally combine forecast models and their inherent uncertainty with incoming noisy observations. The key idea in our work here is to achieve increased forecast capabilities by judiciously combining machine-learning algorithms and data assimilation. We combine the physics-agnostic data-driven approach of random feature maps as a forecast model within an ensemble Kalman filter data assimilation procedure. The machine-learning model is learned sequentially by incorporating incoming noisy observations. We show that the obtained forecast model has remarkably good forecast skill while being computationally cheap once trained. Going beyond the task of forecasting, we show that our method can be used to generate reliable ensembles for probabilistic forecasting as well as to learn effective model closure in multi-scale systems.
\end{abstract}


%
\section{Introduction}
Designing computationally cheap models able to predict a system's state in the future is of utmost importance across disciplines in science and engineering. Often modellers face the situation when the underlying model is unknown and has to be inferred from observational data. There exists a plethora of data-driven approaches but most face the problem that their forecast skill is severely impeded by inevitable contamination of observational noise. This work proposes a method that produces cheap surrogate models from noisy observations and provides remarkably good forecast horizons. Data-driven physics-agnostic modelling has become particularly attractive for modelling high-dimensional 
complex systems where the detailed evolutionary laws are either unknown or are too complex to be resolved numerically, and provide a desirable cost-effective alternative to the numerical simulation of high-dimensional possibly stiff dynamical systems (see, for example, \cite{HanEtAl18,HanEtAl19,RaissiEtAl18,RaissiEtAl19}). Designing such surrogate dynamical models can be formulated abstractly as a problem of function approximation under supervision: mapping a current state to a new state at a later point in time, using information of a given training data set. Being able to approximate this function with satisfactory accuracy then allows for forecasting by mapping unseen data. The concept of supervised learning can further be exploited to find computationally tractable reduced models for a subset of resolved variables. This problem arises in the context of multi-scale systems where one is typically interested in the dynamics of the slow resolved variables. Reducing a stiff potentially high-dimensional multi-scale system to an effective evolution equation for the slow variables only has obvious computational advantages, reducing the dimensionality and allowing for a larger time step in the numerical discretisation. In this context the aim of machine learning is to learn the so called closure term which parametrizes the effect of the fast unresolved degrees of freedom on the resolved slow dynamics. Ideally one aims at determining the closure term when only observations of the resolved variables are available. 

A particularly simple and computationally cheap machine learning technique involves random feature maps \citep{RahimiRecht08}. In the same spirit as expressing a function as a linear combination of basis functions in some Hilbert space such as for example Fourier or Chebychev basis functions, here functions are expressed as linear combinations of functions $\sigma (\mathbf{w}^{\rm T} \mathbf{u} + b)$, the so called {\em{random features}},where $\mathbf{w}$ and $b$ are randomly drawn. It was shown that linear combinations of random feature maps are able to approximate continuous functions arbitrarily close, a property known as the universal approximation property \citep{ParkSandberg91,Cybenko89,Barron93}. We will use the special choice $\sigma(z) = \tanh(z)$ in this paper. The task of training consists of learning the coefficients of the linear combination which can be achieved efficiently by linear ridge regression; see \cite{RahimiRecht08b,Bach17a,Bach17b,SunGilbertTewari19} for recent accounts. In the context of dynamical systems the framework of random feature maps was extended to include internal degrees of freedom with their own dynamics in so called echo-state networks \citep{MaassEtAl02,Jaeger02,JaegerHaas04,PathakEtAl18}. Random feature maps and their extensions have been successfully used when the training data set is noise-free, e.g. when the data originate from precise measurements of the model states. In the case when the training data set instead consists of noisy state observations, we shall see that the standard machine learning approach of linear regression becomes suboptimal and provides a mapping that is less suitable for forecasting unseen data.

Incorporating noisy observations to estimate the state of a dynamical system is a problem addressed by data assimilation (DA) or filtering  \citep{Majda12,LawStuart,ReichCotter}. In DA incoming noisy observations are optimally incorporated to increase the predictability of a given forecast model. DA can be used to estimate the state variables as well as unknown parameters of the model. DA now constitutes a standard tool in science and engineering and is used, for example, in numerical weather forecasting where it is the main driver for the increased predictability enjoyed over the past decades. DA has already been used successfully in data-driven modelling where the forecast model is constructed using Takens' embedding theorem and phase space reconstruction \citep{HamiltonEtAl16}. More recently,
data assimilation techniques have been combined with techniques from machine learning in order to perform combined state and parameter estimation. We mention in particular the work of \cite{ARS18,DSLR19,BBCB19,BBCB20,BCBB20a} on smoothing techniques and the recent work of \cite{BFM20} on sequential learning of both state and parameters. Monte Carlo and optimization based methods 
for combined state and parameter estimation have also already been investigated extensively in the computational statistics community; see \cite{kantas2015} for a review. However, most of these techniques are not applicable to the high-dimensional inference problems that arise from semi-parametric representations of the unknown dynamical propagator maps. 

Here we shall combine the random feature map architecture within a sequential data assimilation procedure. The idea is to learn the coefficients of the linear combination of the random feature map approximation sequentially by updating them in time with incoming observations rather than performing linear regression on the entire training data set. The forecast model within the data assimilation procedure is given by the random feature map model itself. It should be noted, however, that the resulting combined state-parameter estimation problem is no longer linear. Within such a setting, an attractive DA framework is provided by ensemble Kalman filters (EnKF) where the statistical information needed for the Kalman filter is provided through a Monte Carlo approximation of an ensemble of model forecasts \citep{Majda12,LawStuart,ReichCotter}. Such ensemble filters were shown to perform very well for nonlinear dynamical systems unlike the classical Kalman filter. We coin our methodology with the acronym RAFDA, standing for {RA}ndom {F}eature maps and {D}ata {A}ssimilation. We shall consider three prototypical dynamical systems to numerically investigate RAFDA and to illustrate how RAFDA is able to extend the good forecast skills of classical random feature maps to the case of noisy training data. We consider the Lorenz-63 model, the Kuramoto-Sivashinsky equation and the multi-scale Lorenz-96 system to show how RAFDA achieves the goals we set out above. Besides an improved forecast skill of single trajectories compared to classical random feature maps, RAFDA naturally extends to probabilistic forecasting when used in an ensemble setting. We will see that our method generates reliable ensembles for which each ensemble member has equal probability of being closest to the truth. In these applications the aim is to provide a cheap surrogate model which is learned from noisy training data {\em{without}} any information about the underlying dynamical system. In the context of multi-scale systems scientists often know the form of the physical model for the resolved slow scales but lack detailed information about the closure term driving the physical model. The issue of such subgrid-scale parametrisations has recently been addressed anew using machine learning techniques; see for example  \cite{DuebenBauer18,RaspEtAl18,GagneEtAl19,BoltonZanna19,BCBB20b,Nadiga21}. 
For the Lorenz-96 multi-scale system we show that given noisy observations of the slow variables RAFDA can determine the closure term and can be used to perform subgrid-scale parametrization.\\ 

The paper is organised as follows. In Section~\ref{sec:method} we develop our RAFDA methodology. In Section~\ref{sec:L63} we present applications to the Lorenz-63 model, where we test in particular for the dependency of the forecast skill on the length of the training data set, the noise level and the reservoir dimension. We further show that ensembles generated through the combined random feature map and data assimilation procedure provide reliable ensembles to be used in probabilistic ensemble forecasting. We further consider the Kuramoto-Sivashinsky equation as a paradigmatic model for spatio-temporal chaos in partial differential equations in Section~\ref{sec:KS}. In Section~\ref{sec:L96} we consider the multi-scale Lorenz-$96$ model and show  that RAFDA can be used to learn closure models for the effective slow dynamics. We conclude in Section~\ref{sec:discussion} with a discussion and an outlook.

%
\section{Computational methods}
\label{sec:method}
%
Consider a $D$-dimensional dynamical system 
\begin{equation} \label{eq:ODE}
\dot {\u} = \mathcal{\boldsymbol{F}}(\u),
\end{equation}
which is observed at discrete times $t_n = n\Delta t$ of interval length $\Delta t>0$, $n = 0,\ldots,N$, 
to yield a noisy time series of vector-valued observations 
\begin{equation} \label{eq:observations}
\uobs_n = \u_n + \Robs^{1/2}\,  \boldsymbol{\eta}_n
\end{equation}
with $\uobs_n\in\R^{D}$ , $\u_n = \u(t_n)$, measurement error covariance matrix $\Robs\in \R^{D\times D}$ and independent and normally distributed noise 
$\boldsymbol{\eta}_n\in  \R^{D}$, that is, $\boldsymbol{\eta}_n \sim {\mathcal N}({\bf 0},{\bf I})$. We will typically work with
a measurement error covariance matrix 
\begin{equation}
\Robs = \eta {\bf I}
\end{equation}
with scalar variance parameter $\eta \ge 0$. The special case $\eta = 0$ corresponds to exact state observations. In 
this paper, we are primarily interested in noisy state observations, that is, in $\eta >0$.

For the purpose of data-driven modelling and machine learning it is instructive to view the evolution of the time-dependent model state $\u(t)$ in the time interval $\Delta t$  as a propagator map
\begin{equation} \label{eq:propagator1}
\u_{n+1} = \Psi_{\Delta t}(\u_n).
\end{equation}
The aim of data-driven modelling is then to find an approximation of this map and construct a surrogate model
\begin{equation} \label{eq:propagator2}
\hat \u_{n+1} = \Psi_{S}(\hat \u_n)
\end{equation}
with $\hat\u_0 = \u_0$, which is to be learned from the observational data (\ref{eq:observations}). The observational data may come as outputs from a known model or are given by actual observations without any knowledge of the underlying model. In the case when the model is known a surrogate model may still have computational advantages, in particular for high-dimensional models and/or for stiff multi-scale models, where solving the full model may require prohibitively small step-sizes $\delta t\ll \Delta t$. One can also encounter the case when the underlying dynamics (\ref{eq:ODE}) is only partially known. This situation arises typically in multi-scale systems, where $\utruth (t)$ characterises the resolved (slow) degrees of freedom and the analytic form of $\mathcal{\boldsymbol{F}}$ is known but the effect of the unresolved dynamics on the resolved ones has to be inferred from data.

Independent of whether the underlying model (\ref{eq:ODE}) is known or not, one ends up with a combined problem of having to estimate the states $\u_n$  at times $t_n=n\Delta t$ as well as the functional form of the propagator $\Psi_{\rm S}$ from the noisy data $\uobs_n$, $n=0,\ldots,N$. This problem has been well studied in the literature, both within parametric and non-parametric settings, using different approximation tools such as radial basis functions (RBF) and reproducing kernel Hilbert spaces (RKHS). In this paper, we focus on a particular class of shallow neural networks and the RKHS induced by their random feature maps. See, for example, \cite{Bach17a,Bach17b,SunGilbertTewari19}  for recent theoretical results on such approximations.  In particular, the attractiveness of this RKHS, as compared to more complex machine learning architectures such as those considered for example in \cite{QM20}, is its easy embedding into sequential data assimilation via the ensemble Kalman filter \citep{Evensen}, as we will demonstrate in this paper.

We now provide a short derivation of the random feature approximation from a RKHS and Gaussian process perspective
\cite{Bishop,RasmussenWilliams}. The starting point is a kernel function $k(\u,\v)$, which we define to be
\begin{equation}\label{eq:kernel}
k(\u,\v) = \int_{\mathbb{R}^{D+1}} \tanh(\w_{\rm in}^{\rm T}\u + b_{\rm in}) \tanh(\w_{\rm in}^{\rm T}\v + b_{\rm in}) 
\,p(\w_{\rm in}) p(b_{\rm in})\,{\rm d}\w_{\rm in} {\rm d}b_{\rm in}
\end{equation}
with $p(\w_{\rm in})$ some multivariate distribution of dimension $D$ and some univariate 
distribution $p(b_{\rm in})$. We denote the associated reproducing kernel Hilbert space by $\mathcal{H}_k$. 
Monte Carlo integration with 
$D_r$ terms is used to approximate the integral in (\ref{eq:kernel}) to obtain
\begin{equation}\label{eq:kernel_MC}
\widehat{k}(\u,\v) = \frac{1}{D_r} \sum_{k=1}^{D_r} \tanh(\w_{{\rm in},k}^{\rm T}\u + b_{{\rm in},k})
\tanh(\w_{{\rm in},k}^{\rm T}\v + b_{{\rm in},k}) 
\end{equation}
with $\w_{{\rm in},k} \sim p(\w_{\rm in})$ and $b_{{\rm in},k} \sim p(b_{\rm in})$ independently distributed. Using the approximation (\ref{eq:kernel_MC}) as the covariance function of a Gaussian process leads to the following random feature representation 
\begin{equation} \label{eq:r0}
\Psi_S(\u) = \sqrt{\frac{1}{D_r}}\sum_{k=1}^{D_r} w_k  \tanh(\w_{{\rm in},k}^{\rm T}\u + b_{{\rm in},k})
\end{equation}
of realisations from the Gaussian process $\mathcal{G}(0,\widehat{k})$ with the $w_k$'s 
independently and standard Gaussian distributed. Note that
\begin{equation}
\widehat{k}(\u,\v) = \mathbb{E}[\Psi_S(\u)\Psi_S(\v)],
\end{equation}
where the expectation is taken over the random weights $w_k$ in (\ref{eq:r0}). 
The Gaussian process $\mathcal{G}(0,\widehat{k})$ and its random feature 
realisations (\ref{eq:r0}) provide a prior for the estimation of the desired propagator $\Psi_S(\u)$ in (\ref{eq:propagator2})
in case of $D=1$. 

Our random feature approximation to (\ref{eq:propagator2}) for $D\ge 1$ proceeds now as follows: 
We define $D_r$-dimensional feature maps, which we write in vector form as
\begin{align}
\boldsymbol{\phi} (\u) = \tanh(\Win \u + \bin) \in \mathbb{R}^{D_r}
\label{eq:r}
\end{align}
using a weight matrix 
$$
\Win = (\w_{{\rm in},1},\ldots,\w_{{\rm in},D_r})^{\rm T} \in \mathbb{R}^{D_r\times D}
$$ 
and a bias $$
\bin = (b_{{\rm in},1},\ldots,b_{{\rm in},D_r})^{\rm T} \in \mathbb{R}^{D_r \times 1}.
$$ 
The weight matrix and the bias are chosen randomly and independently of the observed $\uobs_n$, $n=0,\ldots,N$,
according to the distributions $p(\w_{\rm in})$ and $p(b_{\rm in})$, respectively. It is pertinent to mention that here the randomly drawn hyper-parameters $\Win$ and $\bin$ are kept fixed once drawn and are not learned. 

The surrogate propagator (\ref{eq:propagator1}) is then defined by
\begin{align}
\Psi_S(\u) = \W \boldsymbol{\phi}(\u),
\label{eq:Wr}
\end{align}
where $\W \in \mathbb{R}^{D\times D_r}$ is a matrix of coefficients which will be learned from the data $\uobs_n$, $n=0,\ldots,N$. 
The case $D=1$ corresponds to 
\begin{equation} \label{eq:Wscaling}
\W = \left( \frac{w_1}{\sqrt{D_r}},\ldots,\frac{w_{D_r}}{\sqrt{D_r}}\right) \in \mathbb{R}^{1\times D_r}
\end{equation}
in (\ref{eq:r0}). The fact that the coefficients $\W$ appear linearly in (\ref{eq:Wr}) allows for learning by simple linear regression and makes random feature maps a computationally very attractive network architecture. Despite its simplicity, it was shown that random feature maps and their associated RKHS enjoy the universal approximation property \citep{ParkSandberg91,Cybenko89,Barron93}, ensuring that their associated RKHS is dense in the space of continuous functions. Their success in practical applications, however, depends crucially on an appropriate choice of two random sets of parameters $\Win$ and $\bin$, that is, the distributions $p(\w_{\rm in})$ and $p(b_{\rm in})$ in (\ref{eq:kernel}). It is also permissible to consider different Gaussian prior distributions for the weights $\W$ leading to different Gaussian process priors. We will comment on these aspects in more detail later.

The Monte Carlo approximation perspective on the random feature maps (\ref{eq:Wr}) suggests that any function $\Psi_{\Delta t} \in
\mathcal{H}_k$ can be approximated by a finite random feature representation (\ref{eq:Wr}) with an error rate of
$\mathcal{O}(1/\sqrt{D_r})$ which does not contain terms exponentially increasing with the dimension $D$ of the underlying dynamical system (see \cite{EEtAl20} for a precise statement). The rate constant, however, may be large depending on the underlying dynamical system and its dimension.

In the next two subsection, we discuss how to employ random feature maps to estimate (\ref{eq:Wr}) from data. 
We first describe the standard linear regression approach which assumes a sequence of exact state observations 
$\uobs_n$, $n\ge 0$. We then present our sequential data assimilation approach to the simultaneous estimation of the model 
states $\u_n$ and the model parameters $\W$.\\

\noindent
{\em{Remark}}: Recently there has been a lot of interest in echo-state networks and reservoir computing \citep{Jaeger02,JaegerHaas04,PathakEtAl18} which involve internal reservoir dynamics. For the examples considered here we find that this additional complexity is not necessary to achieve good forecasting skill. Random feature maps have recently also been extended to solution maps of partial differential equations; see, for example, \citep{NelsenStuart20}. All these contributions assume exact state observations and ignore measurement errors.

%
\subsection{Zero measurement noise: Linear Regression}
%
If the measurement noise in (\ref{eq:observations}) is zero, that is $\utruth_n = \uobs_n$, the external weight matrix $\W \in \mathbb{R}^{D\times D_r}$, which maps the random features to the state variable at the next time step, can be 
determined via linear ridge regression (LR) over a training data set of length $\Ntrain$ \cite{Bishop,EEtAl20,NelsenStuart20}.
More precisely, we seek the minimiser of
 \begin{equation} \label{eq:RR}
\mathcal{L}(\W) = \frac{1}{2} \| \W \Res - \U^{\rm o}\|_{\rm F}^2 + \frac{\beta}{2} \|\W\|^2_{\rm F},
\end{equation}
where $\|\mathbf{A}\|_F$ denotes the Frobenius norm of a matrix $\mathbf{A}$, 
$\U^{\rm o}\in \R^{D\times \Ntrain}$ is the matrix with columns $\uobs_{n}$, $n=1,\ldots,\Ntrain$, and $\Res\in \R^{D_r\times \Ntrain}$ 
consists of columns 
\begin{equation} \label{eq:feature_map_r}
{\boldsymbol{\phi}}_{n} = \boldsymbol{\phi}(\uobs_{n-1}),
\end{equation}
$n=1,\ldots,\Ntrain$ with fixed draws of the the random internal parameters $(\Win,\bin)$ and $\boldsymbol{\phi}(\u)$ given by (\ref{eq:r}). 
The parameter $\beta > 0$ is used for regularization. The minimiser of (\ref{eq:RR}), which we denote by 
$\W_{\rm LR}$, is explicitly given by
\begin{equation}
\W_{\rm LR} = \U^{\rm o}\Res^{\rm T} \left( \Res \Res^{\rm T}+ \beta {\bf I} \right)^{-1} \,.
\label{eq:WLR}
\end{equation}
Algorithm~\ref{algo:LR} below provides a pseudocode outlining the relevant input parameters.

Bounds on the empirical risk
$$
\mathcal{R}_\Ntrain (\W_{\rm LR}) = \frac{1}{2\Ntrain}  \| \W_{\rm LR} \Res - \U^{\rm o}\|_{\rm F}^2 
$$
and its generalization error have been derived in \cite{EEtAl20} for a regularized cost functional slightly different from (\ref{eq:RR}). More precisely, under some additional assumptions on the kernel function $k(\u,\v)$, the authors obtain a rate of $\mathcal{O}(1/D_r + 1/\sqrt{\Ntrain})$ provided the regularization parameter is chosen appropriately. 

From a Gaussian process perspective, the linear ridge regression estimator (\ref{eq:WLR}) can be interpreted as the MAP estimator
to the following Bayesian inference problem \cite{Bishop}: Let us assume $D=1$ for simplicity and that the prior distribution for the desired 
propagator map is provided by the Gaussian process $\Psi_S(u) \sim \mathcal{G}(0,\widehat{k})$. Then the negative log-likelihood 
function of the data is of the form
$$
l(\U^{\rm o}|\Psi_S) = \frac{D_r}{2\beta} \sum_{n=1}^N \| u_{n}-\Psi_S(u_{n-1})\|^2\,,
$$
which corresponds to an additive Gaussian model error $\xi_n$ in (\ref{eq:propagator2}), that is,
\begin{equation} \label{eq:model_error}
\hat u_{n+1} = \Psi_S(\hat u_n) + \xi_n,
\end{equation}
of mean zero and variance $\beta/D_r$. Setting $\beta = D_r$ in (\ref{eq:RR}) corresponds to the standard error model as 
expected from (\ref{eq:Wscaling}) since $\widehat{k}(u_{n-1}^{\rm o},u_{m-1}^{\rm o}) = D_r^{-1} \boldsymbol{\phi}_n^{\rm T} \boldsymbol{\phi}_m$. As in many machine learning applications, the data $\U^{\rm o} \in \mathbb{R}^{1\times N}$ is assumed to be noise-free. We shall explore in Sections~\ref{sec:L63}-\ref{sec:L96} in how 
far linear ridge regression can be used to construct a forecast model trained on noisy observations $\U^{\rm o}$ 
and zero model errors, that
is $\xi_n \equiv 0$ in (\ref{eq:model_error}), instead.

Finding an appropriate weight matrix $\W$ in (\ref{eq:Wr}) has also been investigated from the perspective of 
data-driven approximations of the underlying Koopman operator; the so called extended dynamic mode decomposition 
(EDMD) \citep{WKR15}. The key idea is to first use the complete data set $\Res$
of some observed user-specified feature maps $\boldsymbol{\phi}(\u^{\rm o}_n)$, $n=0,\ldots,N$, which forms the so called {\em{dictionary}}, to construct an approximation of the Koopman operator and to then use its eigenfunctions to approximate the identity map, which in turn implies a computable approximation to the propagator map (\ref{eq:propagator2}) as a linear combination of the features maps, that is, in the form of (\ref{eq:Wr}).
This approach is computationally quite involved as it requires the solution of high-dimensional eigenvalue problems. The problem
of extracting Koopman operators from noisy data $\U^o$ has, for example, been addressed in \cite{HRDC17}. 

We finally mention that the quadratic penalty term in (\ref{eq:RR}) can be replaced by a sparsity enforcing penalty as suggested
in \cite{Brunton3932} and extended to noisy state observations in \cite{KBK20}. It is however not obvious how to extend this, so called,
{\it sparse identification of nonlinear dynamics} (SINDy) approach to an online learning procedure within a Bayesian inference 
framework.

\IncMargin{1em}
\begin{algorithm}
\SetKwData{Left}{left}\SetKwData{This}{this}\SetKwData{Up}{up}
\SetKwFunction{Union}{Union}\SetKwFunction{FindCompress}{FindCompress}
\SetKwInOut{Input}{input}\SetKwInOut{Output}{output}
\SetKwInOut{InputData}{input data}
\SetKwInOut{InputPara}{parameters}
\SetKwInOut{Construct}{construct}
\SetKwInOut{Compute}{compute}
\InputData{time series $\uobs_n$, $n=1,\dots,N$}
\InputPara{regularization parameter $\beta$, reservoir dimension $D_r$,\\ internal parameters $\Win\in \R^{D_r\times D}$, $\bin\in\R^{D_r}$}
\BlankLine
\emph{perform the following:}\\
\Construct{observation matrix $\U^{\rm o} = \left[ \uobs_1,\uobs_2, \cdots , \uobs_N\right]\in \R^{D\times N}$\\
random features $\boldsymbol{\phi}_n = \tanh(\Win \u^{\rm o}_{n-1}+\bin)\in \R^{D_r}$\\
feature matrix $\Res = \left[ \boldsymbol{\phi}_1, \boldsymbol{\phi}_2, \cdots , \boldsymbol{\phi}_{N}\right]\in \R^{D_r\times N}$}
\Output{$\W_{\rm LR} = \U^{\rm o}\Res^{\rm T} \left( \Res \Res^{\rm T}+ \beta {\bf I} \right)^{-1}$}
\caption{Linear Regression (LR)}
\label{algo:LR}
\end{algorithm}\DecMargin{1em}

%
\subsection{Non-zero measurement noise: RAFDA}
%
If the observations are contaminated by noise, we propose here to estimate the weight matrix $\W$ recursively using sequential 
DA \citep{Majda12,LawStuart,ReichCotter}. Whereas the non-recursive estimation described in the previous section using linear ridge regression only utilizes the information contained in the observations, we aim here at optimally estimating the weight matrix using both the observations as well as the underlying dynamical surrogate model (\ref{eq:propagator2}). In particular, we employ a combined state and parameter estimation via state 
augmentation \citep{Majda12,ReichCotter}. More precisely, we formulate the forecast model for constant parameters $\W$ as
\begin{subequations} \label{eq:forecast}
\begin{align} \label{eq:forecast_a}
\u^{\rm f}_{n+1} &= \W_n^{\rm a} \,\boldsymbol{\phi}(\u_n^{\rm a})\\
\W^{\rm f}_{n+1} &= \W^{\rm a}_n,
\end{align}
\end{subequations}
where the superscript f denotes the forecast and the superscript a denotes the analysis defined below. 

Furthermore, the model states $\u_n^{\rm a}$, $\W_n^{\rm a}$ and $\u_{n+1}^{\rm f}$, $\W_{n+1}^{\rm f}$, respectively, are now treated as random variables. To formulate the Kalman filter analysis step, we unravel the weight matrix $\W\in\R^{D\times D_r}$ into a parameter vector 
$\w\in\R^{DD_r}$ with $w_{1:D_r}=W_{11},\dots,W_{1D_r}$, $w_{D_r+1:2D_r}=W_{21},\dots,W_{2D_r}$ and so forth. 
Concatenating further we introduce $\z  = (\u^{\rm T},\w^{\rm T})^{\rm T}
\in \R^{D_z}$ with $D_z = D+D D_r$.  While the forecast step (\ref{eq:forecast}) leads to an update of the random variable 
$\z_n^{\rm a}$ into $\z_{n+1}^{\rm f}$, the analysis step for the mean $\overline \z^{\rm a}_{n+1}$ is provided by
\begin{equation}
\overline \z^{\rm a}_{n+1} = \overline \z^{\rm f}_{n+1} - \K_{n+1} (\H \overline \z^{\rm f}_{n+1}-\uobs_{n+1})
\label{eq:KF1}
\end{equation}
with the observation matrix $\H \in \mathbb{R}^{D\times D_z}$ defined by $\H \z = \u$, i.e., we assume that we observe all state variables $\u$. The Kalman gain matrix $\K$ is given by
\begin{align}
\K_{n+1} = \Pf_{n+1} \H^{\rm T}\left(\H \Pf_{n+1}\H^{\rm T}+\Robs\right)^{-1},
\end{align}
where the forecast covariance matrix is given by
\begin{equation}
\Pf_{n+1} =  \langle \hat \z^{\rm f}_{n+1} \otimes \hat \z_{n+1}^{\rm f} \rangle = 
\left( \begin{array}{cc}
  \langle\hat \u^{\rm f}_{n+1} \otimes \hat\u^{\rm f}_{n+1}\rangle  &  \langle\hat\u^{\rm f}_{n+1}\otimes \hat\w^{\rm f}_{n+1}\rangle\\
  \langle\hat\w^{\rm f}_{n+1}\otimes \hat\u^{\rm f}_{n+1}\rangle  &  \langle\hat\w^{\rm f}_{n+1} \otimes \hat\w^{\rm f}_{n+1}\rangle\\
\end{array} \right).
\label{eq:Pf}
\end{equation}
The angular brackets $\langle f(\z) \rangle$ denote the expectation value of a function $f(\z)$ and the hat denotes the perturbation of $\z_{n+1}^{\rm f}$ from its mean $\overline{\z}_{n+1}^{\rm f} = \langle \z_{n+1}^{\rm f}\rangle$, that is, 
\begin{equation}
\hat \u^{\rm f}_{n+1} = \u_{n+1}^{\rm f} - \overline{\u}_{n+1}^{\rm f}.
\end{equation}
Since we are only observing the state variables $\u$, the Kalman update (\ref{eq:KF1}) can be written to explicitly separate the state and parameter variables as 
\begin{align}
\overline \u^{\rm a}_{n+1} &= \overline \u^{\rm f}_{n+1} -  \Pf_{\u\u} \left( \Pf_{\u\u}+\Robs\right)^{-1} \left(\overline \u_{n+1}^{\rm f}-\uobs_{n+1}\right)\\
\overline \w^{\rm a}_{n+1} &= \overline \w^{\rm f}_{n+1} -  \Pf_{\w\u} \left( \Pf_{\u\u}+\Robs\right)^{-1} \left(\overline \u_{n+1}^{\rm f}-\uobs_{n+1}\right),
\label{eq:KF2}
\end{align}
where $ \Pf_{\u\u} =  \langle\hat \u^{\rm f}_{n+1} \otimes \hat\u^{\rm f}_{n+1}\rangle $ and $\Pf_{\w\u}= 
\langle\hat\w^{\rm f}_{n+1}\otimes \hat\u^{\rm f}_{n+1}\rangle$. 

The Kalman filter is optimal in the sense that $\bar\z^a$ maximizes the likelihood of the observations provided that both the observations and the state variables are Gaussian distributed random variables. Since the forward model (\ref{eq:forecast}) is nonlinear in the augmented state variables and the involved random variables cannot assumed to be Gaussian distributed, the combined forecast and analysis cycle is not well defined and we employ the stochastic EnKF \citep{BurgersEtAl98,Evensen} to define a computationally robust Monte Carlo closure. In particular, define an ensemble of states $\Z \in \mathbb{R}^{D_z\times M}$ consisting of $M$ members $\z^{(i)}\in \mathbb{R}^{D_z}$, $i=1,\ldots,M$, that is,
\begin{equation}
\Z=\left[ \z^{(1)},\z^{(2)},\dots,\z^{(M)} \right],
\end{equation}
with empirical mean
\begin{equation}
\overline{\z} = \frac{1}{M} \sum_{i=1}^M  \z^{(i)},
\end{equation}
and the associated matrix of ensemble deviations
\begin{equation}
\hat \Z=\left[ \z^{(1)}-\overline{\z},\z^{(2)}-\overline{\z},\dots,\z^{(M)}-\overline{\z} \right]
\in \mathbb{R}^{D_z\times M}.
\end{equation}
We define these objects for the forecast and analysis ensembles $\Z_n^{\rm f}$ and $\Z_n^{\rm a}$, respectively, for all $n\ge 0$. Each ensemble member is propagated individually under the forecast step (\ref{eq:forecast}). This defines the update  from the last analysis ensemble $\Z_n^{\rm a}$ to the next forecast ensemble $\Z_{n+1}^{\rm f}$. The EnKF analysis step is now defined as follows. The ensemble deviation matrix $\hat \Z_{n+1}^{\rm f}$ can be used to approximate the forecast covariance matrix (\ref{eq:Pf}) as
\begin{align}
\Pf_{n+1} = 
\frac{1}{M-1}\hat \Z_{n+1}^{\rm f}\,(\hat\Z^{\rm f}_{n+1})^{\rm T}
\in \mathbb{R}^{D_z\times D_z} .
\end{align}
Upon introducing the matrix $\Uobs_{n+1} \in \mathbb{R}^{D\times M}$ of perturbed observations
\begin{equation}
\Uobs_{n+1} = \left[ \uobs_{n+1} - \Robs^{1/2} \boldsymbol{\eta}_{n+1}^{(1)},\uobs_{n+1} - \Robs^{1/2} \boldsymbol{\eta}_{n+1}^{(2)},\ldots,
\uobs_{n+1} - \Robs^{1/2} \boldsymbol{\eta}_{n+1}^{(M)} \right],
\end{equation}
where the $\boldsymbol{\eta}_{n+1}^{(i)} \in \mathbb{R}^D$, $i=1,\ldots,M$, are realisations of independent and normally distributed  random variables (cf (\ref{eq:observations})), 
we obtain the following compact representation of the EnKF update step:
\begin{equation} \label{eq:EnKF}
\Z_{n+1}^{\rm a} = \Z_{n+1}^{\rm f} - \Pf_{n+1}\H^{\rm T}  \left(\H \Pf_{n+1} \H^{\rm T}
+\Robs\right)^{-1} \left(\H\Z^{\rm f}_{n+1}-\Uobs_{n+1}\right).
\end{equation}

The ensemble forecast step defined by (\ref{eq:forecast}), together with the EnKF analysis step (\ref{eq:EnKF}) constitute 
our combined RAndom Features maps and Data Assimilation (RAFDA) method. Training is done with a single long trajectory of length $N$, and the ensemble mean of the weight matrix 
$\W_N^{\rm a}$ at final training time $t_N = \Delta t N$ is denoted by $\W_{\rm RAFDA}$. 
We note that random perturbations, such as in (\ref{eq:model_error}) can easily be included into the forecast step (\ref{eq:forecast_a}). 
Throughout this paper we assume however that the models are deterministic as in
(\ref{eq:propagator1}). In the subsequent two subsections we discuss further algorithmic details of our method.

%
\subsection{Choice of random feature maps}
%
The choice of the random coefficients $\Win$ and $\bin$ in the feature maps (\ref{eq:r}) and their dimension $D_r$ is crucial for the success of our method. In this study, we choose these entries to be independent and uniformly distributed with
\begin{align}
(\Win)_{ij} \sim \mathcal{U}[-w,w] \qquad {\rm{and}} \qquad (\bin)_i \sim \mathcal{U}[-b,b] .
\label{eq:wb}
\end{align}
The hyper-parameters $w>0$ and $b>0$ should be chosen such that the random feature maps (\ref{eq:feature_map_r}) evaluated at the observed data points  cover their full range of $[-1,1]^{D}$ as uniformly as possible, in particularly sampling their nonlinear domain. See the numerical experiment section for more details.
 
We mention that one could also dynamically adapt the internal parameters $(\Win,\bin)$. This would extend our method from random feature maps to two-layer networks, going from RKHS to Barron spaces \citep{ChizatBach18,MeiEtAl18,RotskoffVandenEijnden18,SirignanoSpiliopoulos20,EEtAl19}. We leave this option in combination with data assimilation as a topic for further research.

%
\subsection{Further algorithmic details of RAFDA}
%
The required ensemble size $M$ of the standard implementation (\ref{eq:EnKF}) of an EnKF needs to be of the order 
$\mathcal{O}(D_z)$, which can become prohibitive for typical reservoir dimensions of $D_r \sim\mathcal{O}(10^3)$ 
and potentially high-dimensional dynamical systems. Within the EnKF community further approximations such as localisation 
and inflation have been developed to deal with finite-size effects \citep{Evensen,ReichCotter,HoutekamerZhang16}. Here we follow the concept of B-localisation \citep{HoutekamerMitchell98,Houtekamer01,Hamill01}, where the empirical covariance matrix $\Pf_{n+1}$
is tempered by a symmetric positive definite localisation matrix ${\bf B}$ via the Kronecker product, that is,
\begin{equation}
\widetilde \P^{\rm f}_{n+1} = {\bf B} \circ \Pf_{n+1}.
\end{equation}
The EnKF update step (\ref{eq:EnKF}) is then replaced by
\begin{equation}
\label{eq:EnKF_loc}
\Z_{n+1}^{\rm a} = \Z_{n+1}^{\rm f} - \widetilde \P^{\rm f}_{n+1} \H^{\rm T} \left( \H \widetilde \P^{\rm f}_{n+1} \H^{\rm T}
+\Robs\right)^{-1} \left(\H\Z^{\rm f}_{n+1}-\Uobs_{n+1}\right).
\end{equation}
Localisation implies that certain correlations between variables get reduced or even set to zero if the corresponding entry in
$\mathbf{B}$ is less than one or zero, respectively. This allows to control spurious correlations between uncorrelated variables of $\mathcal{O}(1/\sqrt{M})$, caused by the finite ensemble size. In our numerical experiments we employ the following form of localisation.
The $j$th row of the weight matrix $\W$ is responsible for updating the $j$th component of the dynamic state variable $\u$ under
the forward model (\ref{eq:forecast_a}). We reflect this property in the approximate covariance matrix $\widetilde{\P}_{n+1}^{\rm f}$
by ignoring all correlations between the $j$th row of $\W$ and all components of the observed $\uobs_{n+1}$ except for its
$j$th entry. We still require $M$ to be $\mathcal{O}(D_r)$ to mitigate against possible spurious correlations. The issue of localisation
in the context of combined state and parameter estimation has also been addressed recently in \cite{BFM20}.\\

We also employ multiplicative inflation in which the ensemble members $\z_{n+1}^{(i)}$ 
of the forecast ensemble $\Z^{\rm f}_{n+1}$ are replaced by
\begin{equation}
\z_{n+1}^{(i)} \rightarrow \overline{\z}_{n+1} + \alpha (\z_{n+1}^{(i)}-\overline{\z}_{n+1}),
\end{equation}
$i=1,\ldots,M$, prior to the EnKF data assimilation step (\ref{eq:EnKF})
with $\alpha>1$ being the inflation factor \citep{AndersonAnderson99}. Note that the inflation step maintains the
ensemble mean while increasing the covariance matrix $\Pf_{n+1}$ of the ensemble.\\

We finally need to specify the distribution of the extended state variable $\z$ at initial time $t=0$ from which to draw our initial ensemble. We treat the two components $\u_0$ and $\w_0$ of $\z_0$ as independent, and set  
\begin{equation}
\u_0 \sim {\mathcal N}(\uobs_0,\Robs)
\label{eq:u0}
\end{equation}
and
\begin{equation}
\w_0 \sim {\mathcal N}(\w_{\rm LR},\gamma \mathbf{I}),
\label{eq:w0}
\end{equation}
where $\w_{\rm LR}$ is the vectorial form of the solution $\W_{\rm LR}$ to the ridge regression formulation (\ref{eq:RR})
and $\gamma >0$ is a parameter. Alternatively, one can also use $\w_0 \sim {\mathcal N}({\bf 0},\gamma \mathbf{I})$ if
$\w_{\rm LR}$ is unavailable.

We recall that the linear regression formulation (\ref{eq:RR}) ignores measurement errors in the states, that is, assumes 
$\u_0 = \uobs_0$. The regularization term in (\ref{eq:RR}) corresponds to the choice of prior for $\w_0$ in (\ref{eq:w0}). 
However, we re-emphasize that LR and RAFDA solve different estimation problems.

We provide a pseudocode outlining the relevant input parameters in Algorithm~\ref{algo:RAFDA}. In the subsequent sections we illustrate the proposed RAFDA methodology in several dynamical systems, demonstrate its improved forecast skill compared to the linear regression approach (\ref{eq:WLR}), and investigate the dependency of RAFDA on the noise strength and the available training size $\Ntrain$, as well as showing the impact on the approximation properties of RAFDA of the various choices a modeller needs to take such as the reservoir dimension $D_r$ and  the choice of the random elements $\W_{\rm in}$ and $\b_{\rm in}$.

\IncMargin{1em}
\begin{algorithm}
\SetKwData{Left}{left}\SetKwData{This}{this}\SetKwData{Up}{up}
\SetKwFunction{Union}{Union}\SetKwFunction{FindCompress}{FindCompress}
\SetKwInOut{Input}{input}\SetKwInOut{Output}{output}
\SetKwInOut{InputData}{input data}
\SetKwInOut{InputPara}{parameters}
\SetKwInOut{Construct}{construct}
\SetKwInOut{Compute}{compute}

\InputData{time series $\uobs_n$, $n=0,\dots,N$}
\InputPara{random feature maps: dimension $D_r$, internal parameters $\Win\in \R^{D_r\times D}$, $\bin\in\R^{D_r}$\\
EnKF: ensemble size $M$, observational noise covariance $\Robs$, inflation $\alpha$,\\ $\qquad\quad$ initial ensemble parameters $(\w_{\rm LR},\gamma)$}
\BlankLine
\emph{perform the following:}\\
\emph{initializing ensemble:\\
$\quad\,$  $\Z_0^{\rm a}$ with  members drawn according to $\u_0 \sim {\mathcal N}(\uobs_0,\Robs)$ and $\w_0 \sim {\mathcal N}(\w_{\rm LR},\gamma \mathbf{I})$}\;
\For{$n=1 :  N$}{
 \emph{forecast $\Z_{n-1}^{\rm a} \to \Z_{n}^{\rm f}$:} each ensemble member is propagated according to\\
 $\qquad$           $\u^{\rm f}_{n} = \W_{n-1}^{\rm a} \,\phi(\u^{\rm a}_{n-1})$\;
 $\qquad$           $\W^{\rm f}_{n} = \W^{\rm a}_{n-1}$\;
  \emph{data assimilation analysis update $\Z_{n}^{\rm f} \to \Z_{n}^{\rm a}$:}\\
$\qquad$ inflation \& localisation: $\Pf_n\leftarrow {\bf B} \circ \Pf_n$\\
$\qquad$  $\Z_{n}^{\rm a} = \Z_{n}^{\rm f} - \Pf_{n}\H^{\rm T}  \left(\H \Pf_{n} \H^{\rm T}
+\Robs\right)^{-1} \left(\H\Z^{\rm f}_{n}-\Uobs_{n}\right)$\;
}
\Output{$\W_{\rm RAFDA} = {\text{ensemble average of }} \W^{\rm{a}}_N$}
\caption{Random Feature Map DataAssimilation (RAFDA)}
\label{algo:RAFDA}
\end{algorithm}\DecMargin{1em}


\section{Ordinary differential equations: Lorenz-63 equation}
\label{sec:L63}
We consider as a first test bed the Lorenz-63 system \citep{Lorenz63}
\begin{align}
\dot y &= 10(y-x)\nonumber \\
\dot x &= 28 x - y -xz\nonumber \\
\dot z &= -\frac{8}{3}z + xy
\label{eq:L63}
\end{align}
with $\u=(x,y,z)^{\rm T} \in \mathbb{R}^3$. 
Observations are taken every $\Delta t = 0.02$ time units and $\u_n$ corresponds to the solution at $t_n = n\Delta t$. 
In all simulations a transient of $40$ model time units is discarded to ensure that the dynamics has settled on the attractor. 

To assess the propensity for random feature maps with data assimilation and for classical random feature maps with linear regression to be used as a forecast model we test the forecast models of LR
\begin{align}
\u_{n+1}=\W_{\rm LR}\, \boldsymbol{\phi}(\u_n)
\label{eq:LR}
\end{align}
with $\W_{\rm LR}$ given by (\ref{eq:WLR}), and of RAFDA
\begin{align}
\u_{n+1}=\W_{\rm RAFDA} \,\boldsymbol{\phi}(\u_n),
\label{eq:RAFDA}
\end{align}
where $\W_{\rm RAFDA}$ is determined via the training data as the final outcome of the iterative DA procedure given by (\ref{eq:forecast}) and (\ref{eq:EnKF}). In both cases the reservoir variables are constructed using (\ref{eq:r}). For LR we employ a regularization parameter of $\beta=4\times10^{-5}$ (unless stated otherwise). For RAFDA, unless stated otherwise, we do not use inflation ($\alpha=0$) and set $D_r=M=300$ and $N=4,000$. We assume that we know the observational noise strength $\eta=0.2$.\\
We test the respective forecast capabilities in a validation data set $\u_{\rm valid} (t_n)$, $n\ge 0$. As a quantitative diagnostic for the forecast skill we record the forecast time $\tauf$, defined as the largest time such that the relative forecast error $\mathcal{E}(t_n)=||\u_{\rm valid}(t_n)-\u_n ||^2/||\u_{\rm valid}(t_n)||^2\le \theta$. We choose here $\theta=0.05$. This threshold was chosen to correspond with an eye-ball metric that the truth and the surrogate model have diverged. 

We measure time in units of the Lyapunov time $t\lambda_{\rm{max}}$ with the maximal Lyapunov exponent $\lambda_{\rm{max}}=0.91$. We shall present results for the mean behaviour over realisations, which differ in observations used for the training, in the validation data set, each of which with independently drawn initial conditions, as well as in the random draws of the internal parameters $(\Win,\bin)$ and of the initial ensembles for RAFDA (unless stated otherwise). The different training and validation data sets are generated by drawing random initial conditions, which then each subsequently evolve into random dynamic states on the attractor after a transient time of $40$ model time units. In the following we discuss the effect of the various parameters as well as on the length of the available data and the noise levels on the forecast capabilities of the two models. \\


\subsection{Dependency on the internal parameters}
\begin{figure}[htbp]
\centering
\includegraphics[width = 0.49\columnwidth]{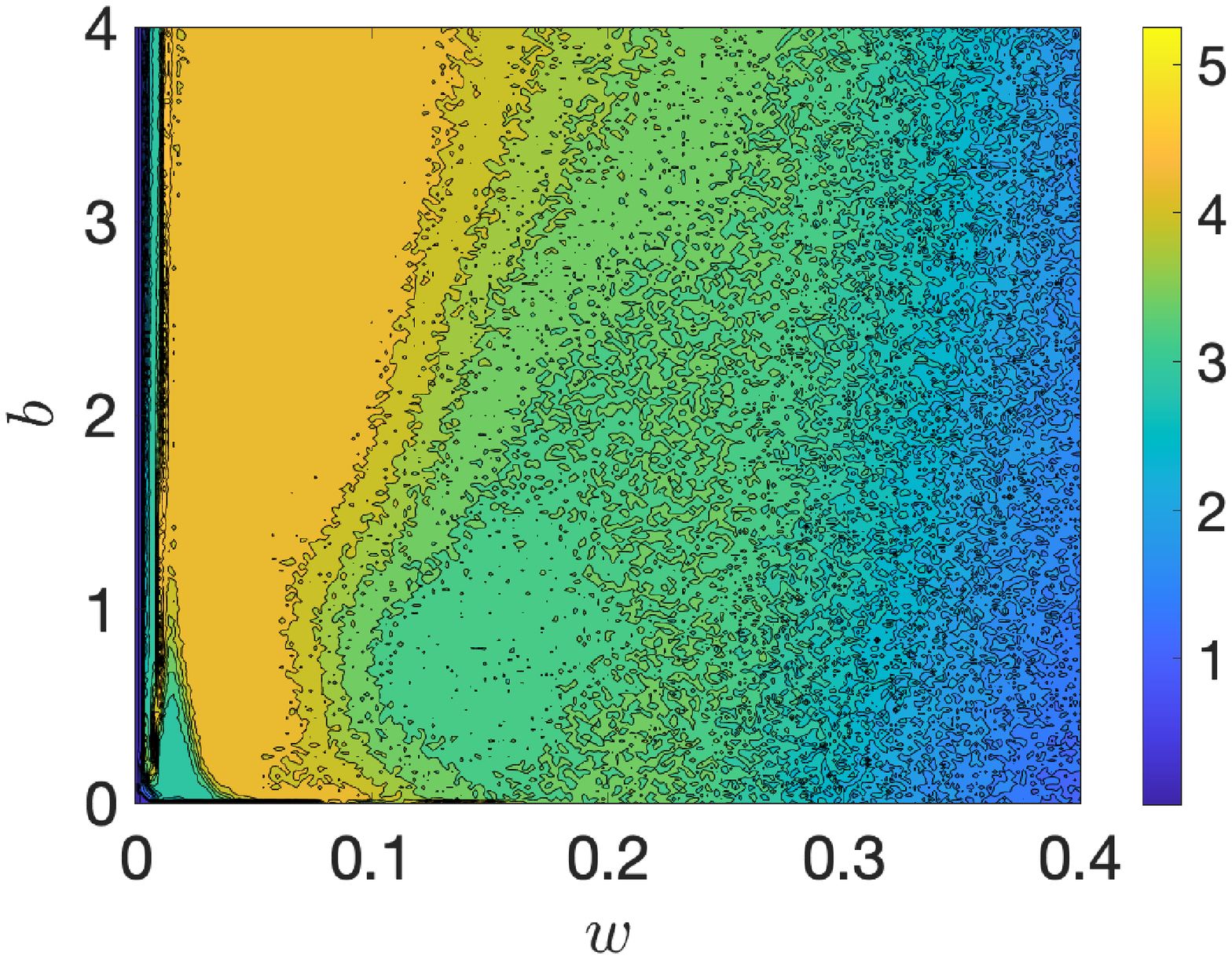}
\includegraphics[width = 0.49\columnwidth]{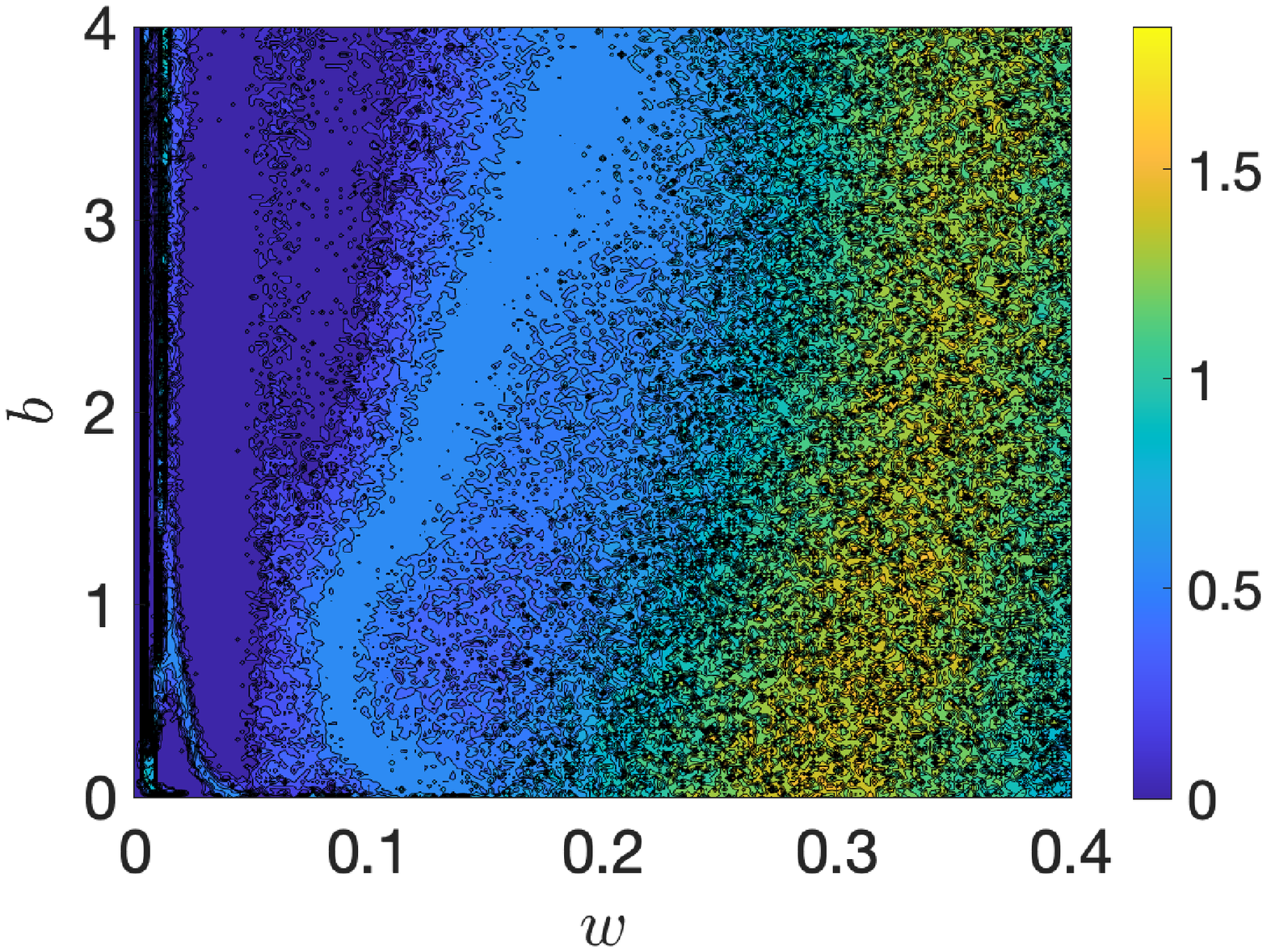}\\
\includegraphics[width = 0.49\columnwidth]{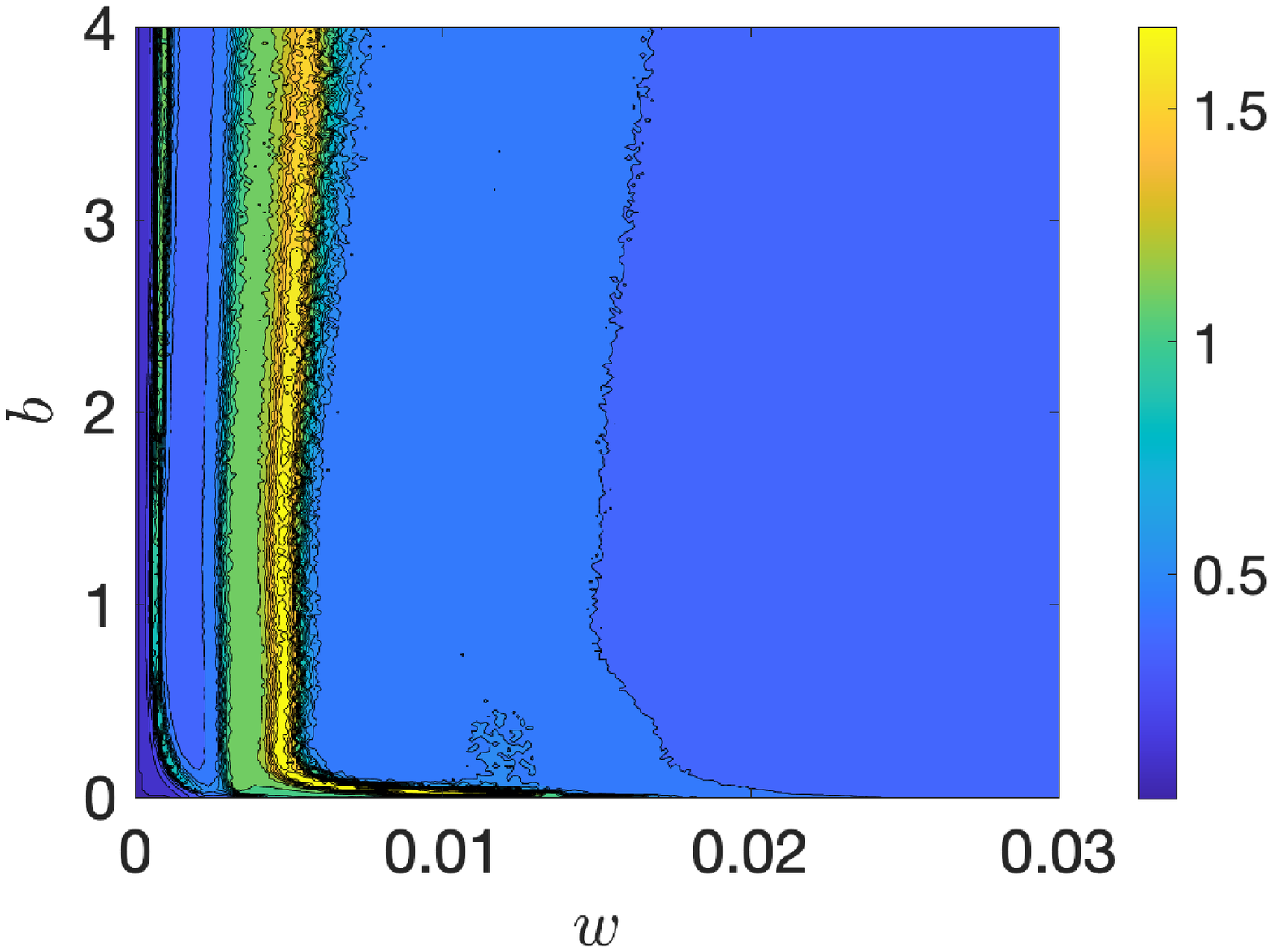}
\includegraphics[width = 0.49\columnwidth]{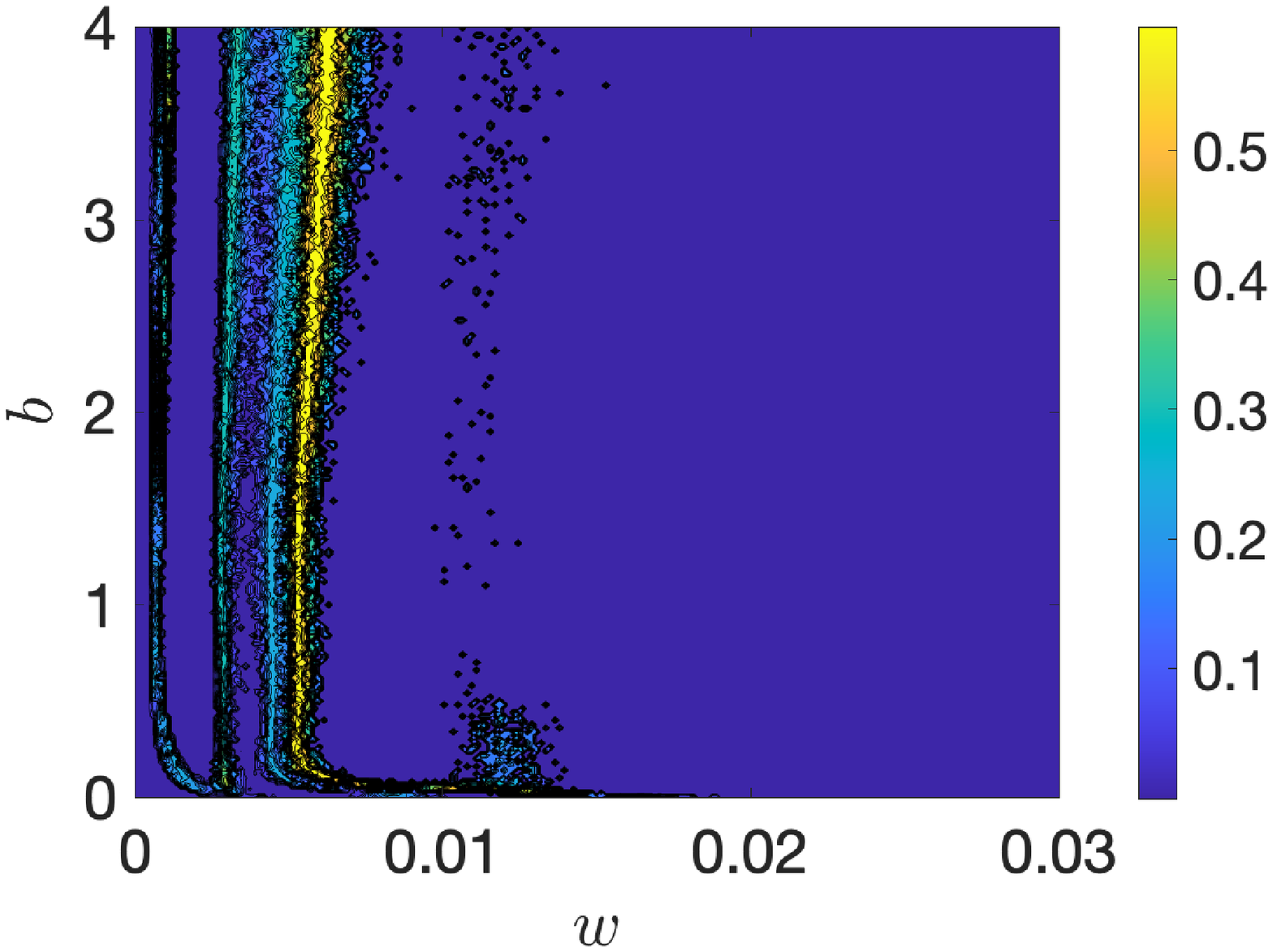}\\
\caption{Contourplot of the forecast time $\tauf$ (left) for different choices of the internal parameters, averaged over $50$ realisations, and the associated standard deviations (right) computed using LR.   
Top: noiseless observations with $\eta=0$. Bottom: noisy observations with $\eta=0.2$.}
\label{fig.L63_paracont}
\end{figure}

\begin{figure}[htbp]
\centering
\includegraphics[width = 0.49\columnwidth]{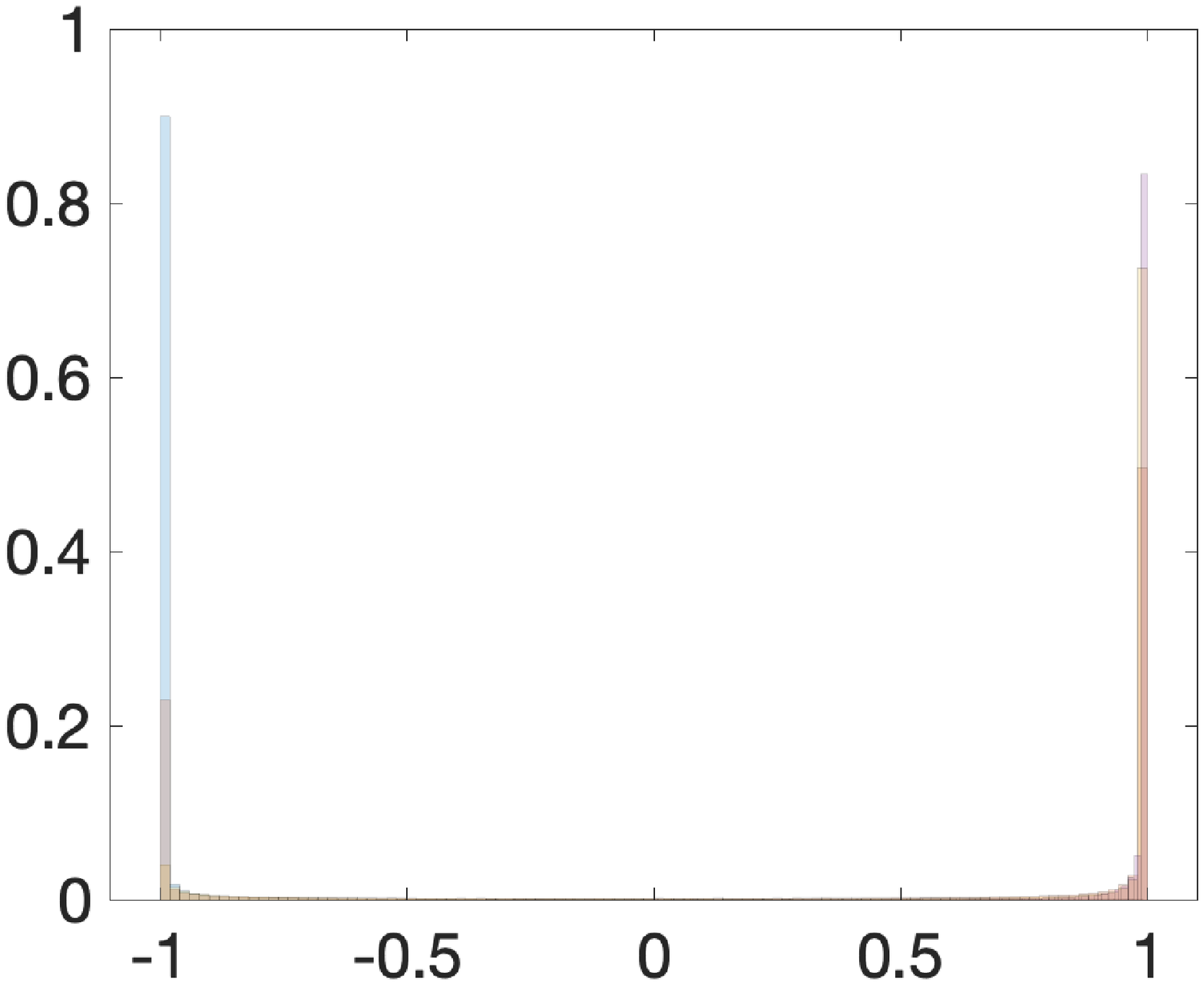}
\,
\includegraphics[width = 0.49\columnwidth]{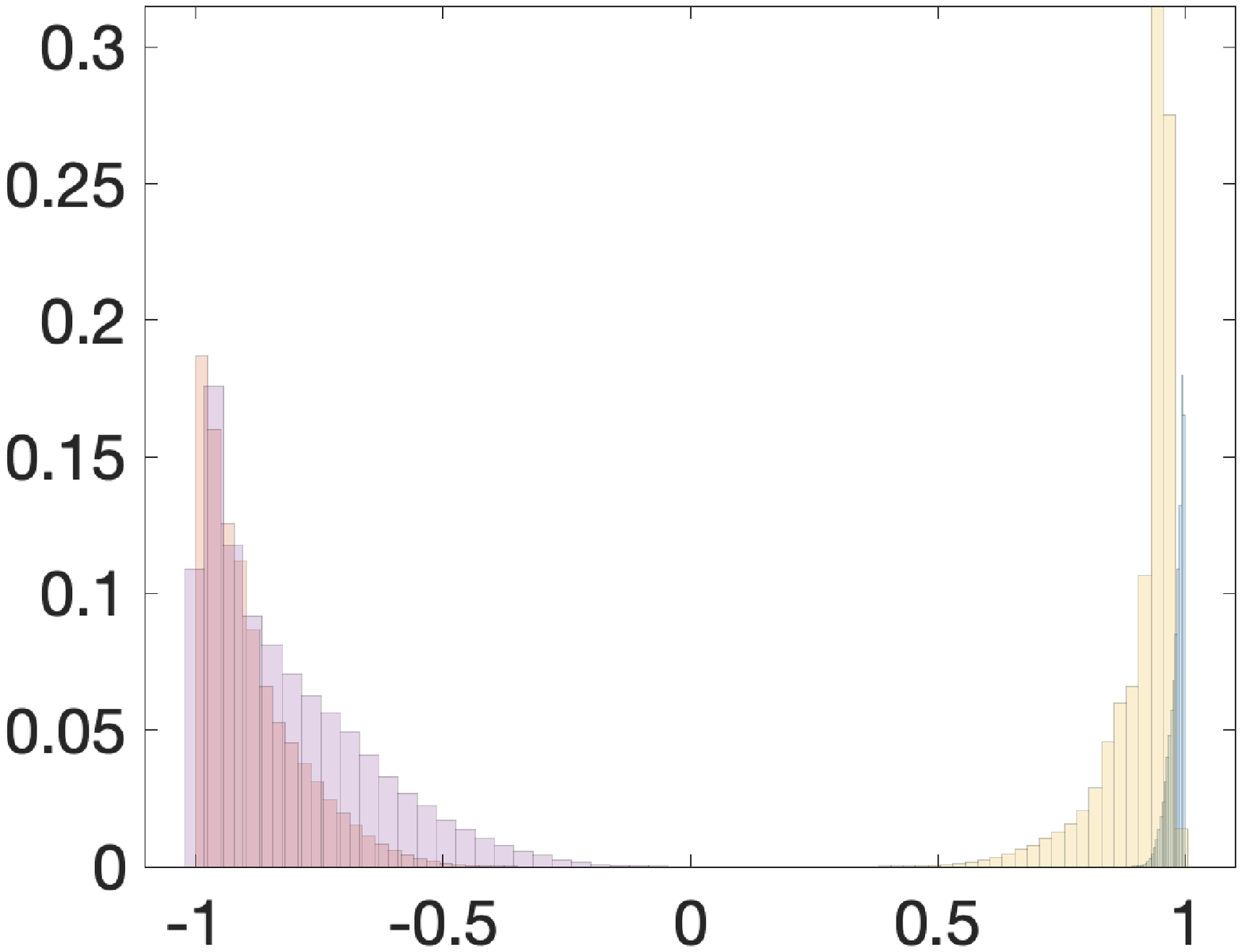}
\caption{Normalized empirical histogram of the random feature map (\ref{eq:r}) $\phi_1$ for four choices of internal parameters $(\Win,\bin)$, drawn for four separate values of $(w,b)$. Left: for internal parameters corresponding to forecast times $\tauf<0.5$. Right: for internal parameters corresponding to forecast times $\tauf>6$.}
\label{fig.L63_para_hist}
\end{figure}

We first address the issue of the choice of the randomly chosen internal parameters $\Win$ and $\bin$. The entries of the internal weight matrix and bias are chosen uniformly randomly over intervals $[-w,w]$ in case of $\Win$ and $[-b,b]$ in case of $\bin$ (cf (\ref{eq:wb})). Figure~\ref{fig.L63_paracont} shows the forecast times $\tauf$ and the standard variations as a function of the two parameters $w$ and $b$ averaged over $50$ independent realisations of (\ref{eq:wb}) using the same training data set and the same validation data set $\uvalid$ but differing in the random draws $(\Win,\bin)$.  The forecast times are estimated using the ridge linear regression matrix $\W_{\rm LR}$ with $N=20,000$ training data points. Results are shown for training on noiseless and on noisy observations with $\eta=0.2$. The forecast capabilities clearly depend sensitively on the choice of $w$ and $b$. In particular, parameter choices leading to excellent forecast times in the noiseless case may not lead to good forecast times in the case of non-zero measurement noise (note the different range in $w$ for the contour plots shown in Figure~\ref{fig.L63_paracont}). The optimal forecast times $\tau_f$ for noiseless data is $5.5$ whereas it drops to $1.8$ when trained on noisy data, illustrating that random feature maps have difficulties when trained on noisy data. Moreover, in the noiseless case parameter combinations $(w,b)$ associated with large forecast times exhibit small variances which renders random feature maps and linear regression insensitive to the particular random draw. In the case of noisy data, however, it is seen that parameter combinations $(w,b)$ corresponding to large forecast times also have larger standard variance, rendering this method unreliable for noisy data. To explore what constitutes a good set of internal parameters corresponding to long forecast times we show in Figure~\ref{fig.L63_para_hist} normalized histograms of one of the components of the random feature map $\phi_j$ (cf (\ref{eq:r})). We pick values $(w,b)$ which we identified to lead (on average) to poor forecast times $\tauf<0.5$ and which we identified to lead to good forecast times with $\tau_f>4$ in the case of noiseless observations (cf. Figure~\ref{fig.L63_paracont}). We then choose randomly drawn typical realisations corresponding to each of the $(w,b)$. The histogram is then generated by evaluating $\phi_1$ along a time trajectory $\u(t)$ of the Lorenz-63 system of length $N=200,000$ sampled every $0.02$ time steps, to explore the range of the random feature map $\phi_1$ for the Lorenz-63 dynamics. For internal parameter choices corresponding to poor forecast times $\tauf<0.5$ the histograms exhibit strongly localised behaviour near $\pm 1$, caused by the $\tanh$-random feature map (\ref{eq:r}) having scale parameters $w$ such that the $\tanh$-function cannot resolve the whole range of dynamical states of the underlying observations. Internal parameters corresponding to long forecast times on the other hand resolve the whole dynamical range within the nonlinear range of the $\tanh$-function. In the following the internal weight matrix and bias are chosen randomly with $w=0.005$ and $b=4$ which yields (on average) large forecast times $\tauf$ for $\eta=0$ and $\eta=0.2$ (cf. Figure~\ref{fig.L63_paracont}).\\

%


\subsection{Dependency of LR on the regularization parameter $\beta$}
\begin{figure}[htbp]
\centering
\includegraphics[width = 0.5\columnwidth]{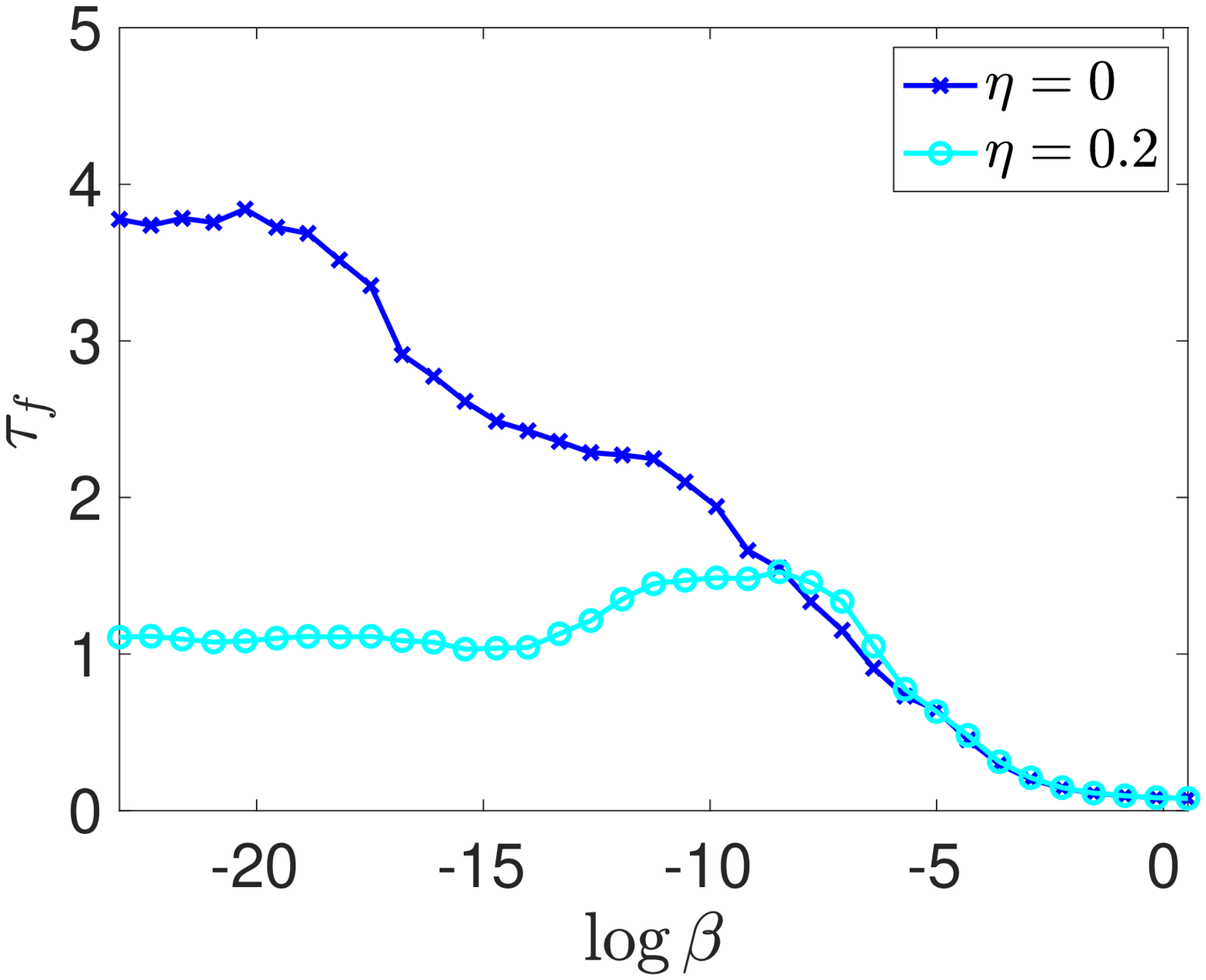}
\caption{Mean of the forecast time $\tauf$ in units of the Lyapunov time for varying regularization parameters $\beta$ for LR for noiseless ($\eta=0$) and for noise contaminated observations ($\eta=0.2$).}
\label{fig.L63_beta}
\end{figure}

Figure~\ref{fig.L63_beta} shows the mean of the forecast time $\tau_f$ obtained from LR for reservoir dimension $D_r=300$ for different choices of the regularization parameter $\beta$, averaged over $500$ realisations, each trained on $N=4,000$ observations. Each realisation uses a different random draw of the internal parameters $\Win$ and $\bin$ as well as different training and validation data sets. 
We observe that when noiseless data are used to train LR, the forecast time plateaus to its maximal value of around $\tau_f=3.8$ for $\log\beta <-20$  (we use the natural logarithm throughout the paper). For noisy observations with $\eta=0.2$ the mean forecast time $\tauf$ is relatively robust to changes in the regularization parameter for a wide range of $\beta$ with $\log \beta \in[-25,-7]$. Within this range the forecast times $\tau_f$ are consistently larger for the training set consisting of noiseless observations.  In the following we employ a regularization parameter of $\beta=4\times 10^{-5}$ ($\log \beta \approx -10$).

%


\subsection{Dependency of RAFDA on the initial ensemble}
\begin{figure}[htbp]
\centering
\includegraphics[width = 0.5\columnwidth]{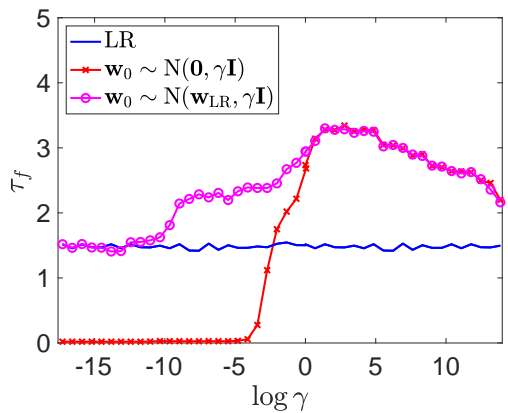}
\caption{Mean of the forecast time $\tauf$ in units of the Lyapunov time for varying  $\gamma$ for RAFDA for noise contaminated observations with $\eta=0.2$. Shown are results when the initial ensemble is drawn around the prior provided by LR $\w_0 \sim {\mathcal N}(\w_{\rm LR},\gamma \mathbf{I})$ (\ref{eq:w0}) and for the unbiased choice $\w_0 \sim {\mathcal N}({\bf 0},\gamma \mathbf{I})$.} 
\label{fig.L63_gamma}
\end{figure}

Turning to RAFDA, we now investigate the influence of the choice of the initial ensemble for the parameters on its performance. We consider initial ensembles drawn in an unbiased fashion $\w_0 \sim {\mathcal N}({\bf 0},\gamma \mathbf{I})$ as well as drawn around the prior provided by LR with $\w_0 \sim {\mathcal N}(\w_{\rm LR},\gamma \mathbf{I})$ as in (\ref{eq:w0}). 
Recall that the small-cap $\w$ denotes the vectorial representation of the parameter matrix large-cap $\W$ which is learned in RAFDA. The model is trained on $N=4,000$ noisy observations with noise level $\eta=0.2$ and a reservoir of dimension $D_r=300$ was used. We chose $M=300$ ensemble members for RAFDA, and averaged over $500$ realisations, each involving independently drawn random internal parameters $\Win$ and $\bin$, independently drawn initial ensembles for RAFDA as well as different training and validation data sets. Figure~\ref{fig.L63_gamma} shows the dependency of the mean forecast time on the initial spread of the ensemble $\gamma$. The forecast time vacillates around $\tau_f\approx 1.5$ for LR (the small fluctuations occur due to each realisation having different independent training and validation data sets). For both types of initial RAFDA ensembles there is a wide range of initial ensemble spreads $\gamma$ for which good forecast times are obtained. When $\gamma$ is too small and the spread of the initial ensemble in the data assimilation component of RAFDA is too small, EnKF experiences filter divergence and collapses, implying a small error covariance matrix $\Pf$. This leads to the filter trusting its own forecast, ignoring the information from incoming observations. The filter then amounts to running the bare  forecast model, leading to poor forecast skills with small $\tau_f \to 0$ for $\gamma\to 0$. Centering the initial ensemble around the LR prior $\w_{\rm LR}$ significantly delays this filter collapse with forecast times $\tau_f\approx 2.5$ when the unbiased initial ensemble has already collapsed near $\log \gamma\approx -4$. When further decreasing the spread of the initial parameter ensemble $\gamma$ with $\log \gamma \lessapprox -10$, the parameters $\w$ of RAFDA, when initialised on the prior provided by LR, are not corrected by the incoming observations and RAFDA and LR perform equally well. 

For large values of $\gamma$, as to be expected, the same forecast times $\tauf$ are produced independent of the chosen initial RAFDA ensemble. The same  forecast times $\tauf$ are obtained for ensembles centred centred around the prior $\W_{\rm LR}$ and for unbiased ensembles for $\log \gamma \approx  0$. For $\log \gamma <0$, centering the initial ensemble around the prior provided by LR produces larger forecast times than LR for noisy data, and generally leads to more robust forecast times with respect to the initial spread $\gamma$ than the unbiased initial ensemble. We remark that the differences between the two initialisation choices depend on the noise level of the training data set; for small noise levels, centering the initial ensemble around the prior provided by LR leads to better forecast times compared to the unbiased initial ensemble for a wider range of values of $\gamma$. In the following we will use initial ensembles centred around the prior provided by LR with $\gamma =1,000$. 

Whereas the particular draw of the initial parameters $(\Win,\bin)$ can have a large impact on the performance of RAFDA (see also Figure~\ref{fig.L63_Dr_hist} below), the DA component of RAFDA seems insensitive to the particular random draw of $\w_0$. For initial ensembles (\ref{eq:u0})--(\ref{eq:w0}), which achieve good analysis error of the state variables, we observed that the forecast times do not depend strongly on the particular random draw.

%


\subsection{Distribution of the learned parameters $W_{\rm{LR}}$ and $W_{\rm{RAFDA}}$}
\label{sec:P0}
\begin{figure}[htbp]
\centering
\includegraphics[width = 0.48\columnwidth]{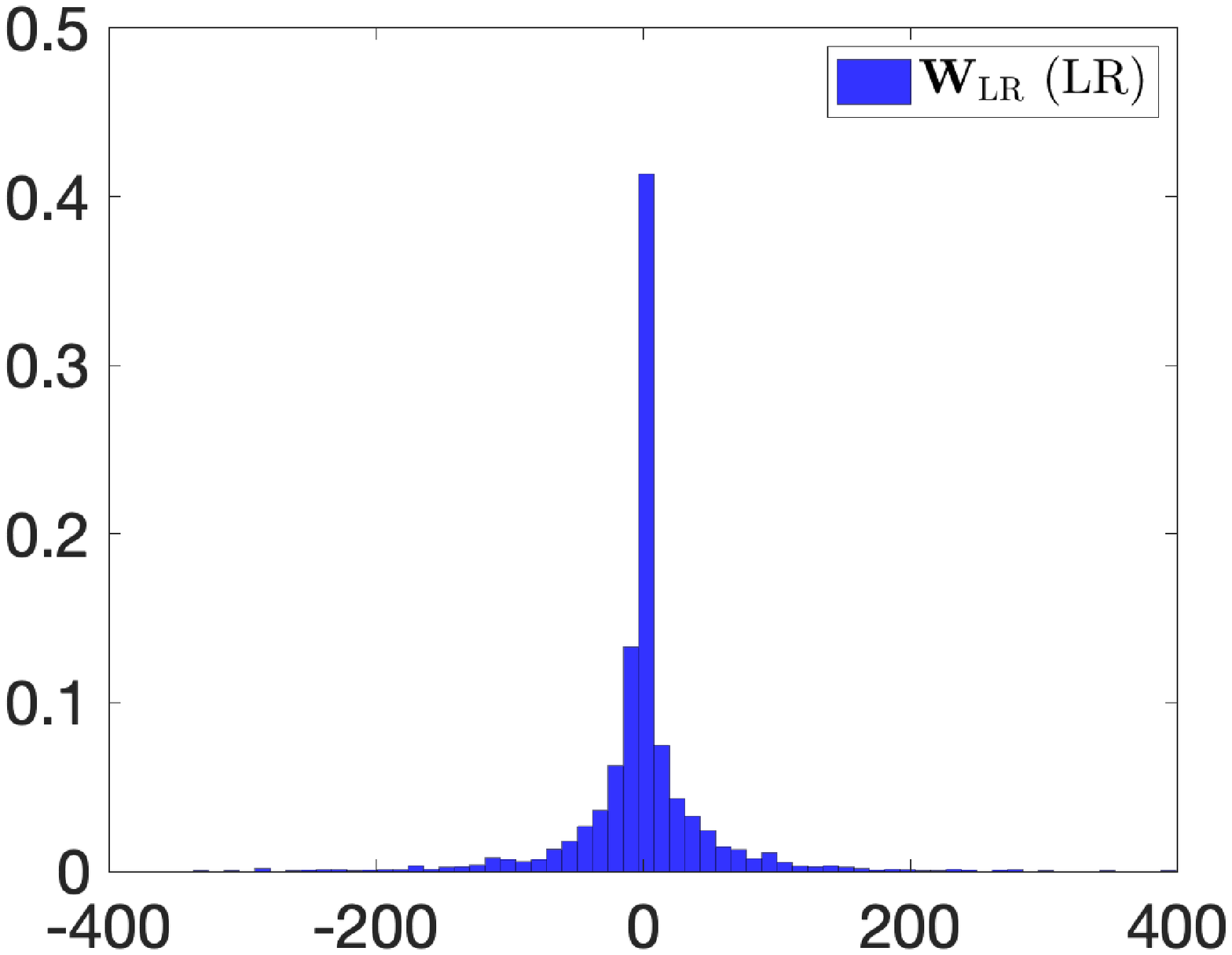}
\includegraphics[width = 0.48\columnwidth]{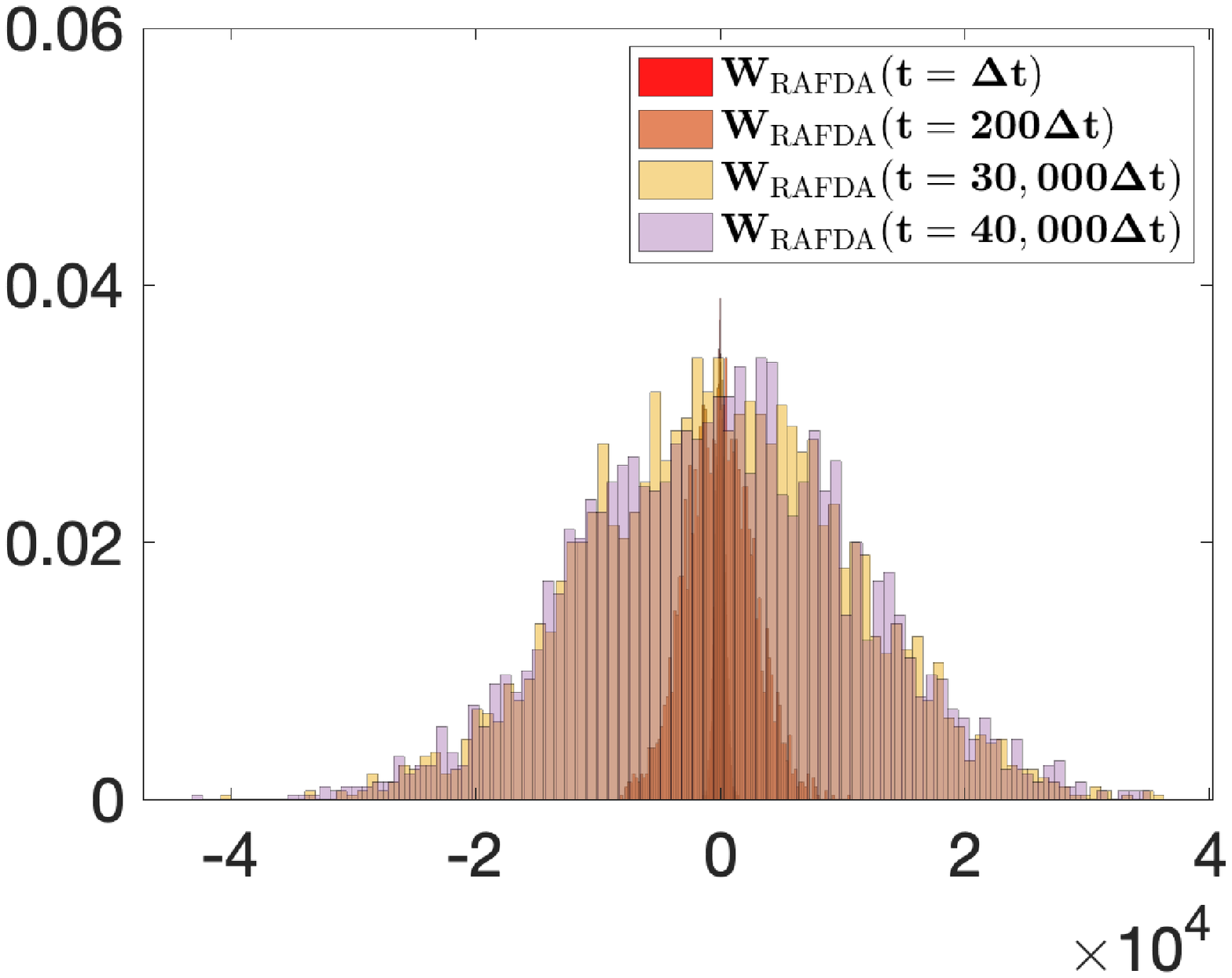}
\caption{Empirical histogram of the learned entries of the weight matrix $\W_{\rm LR}$ (left) and $\W_{\rm RAFDA}$ (right).  For RAFDA several histograms are shown at different times to illustrate the convergence of the distribution during the assimilation procedure.}
\label{fig.L63_Pconv}
\end{figure}

Figure~\ref{fig.L63_Pconv} shows how the learned parameters $\W_{\rm{LR}}$ and $\W_{\rm{RAFDA}}$ are distributed. Shown are the empirical histograms $\W_{\rm{LR}}$ and $\W_{\rm{RAFDA}}$ histogram for a single realisation of a training data set with $N=40,000$, $D_r=M=1,000$ and observational noise covariance $\eta=0.2$. LR yields a forecast time of $\tau_f=0.9$ and RAFDA a forecast time of $\tau_f=4.4$. The parameters $\W_{\rm LR}$ are determined by learning all the data at once, with a distribution which is strongly peaked at small values. The parameters $\W_{\rm RAFDA}$ on the other hand are learned sequentially. Surprisingly, we do not see that individual components of $\W$ (or its vectorial form $\w$) converge, but they converge in distribution. In particular, it is seen that their distribution is near-Gaussian with comparatively much wider support than those obtained by LR. The lack of convergence of the individual entries of $\W$ in RAFDA during the assimilation, we believe, is due to the non-uniqueness of the map $\u_{n+1}=\W\boldsymbol{\phi}(\u_n)$, i.e., the same mapping can be achieved (within the error tolerance of the observational noise) for more than one parameter combinations in $\W$.\\

\subsection{Dependency on the reservoir dimension $D_r$}
\label{sec:Dr}
\begin{figure}[htbp]
\centering
\includegraphics[width = 0.5\columnwidth]{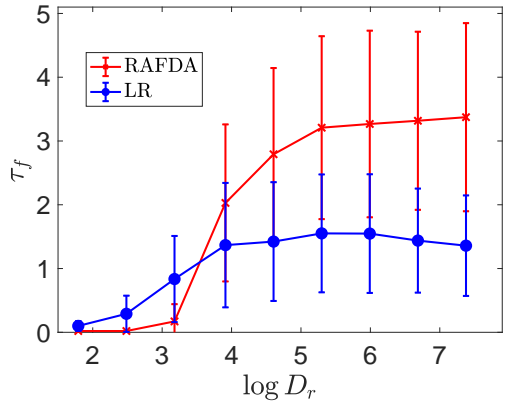}
\caption{Mean of the forecast time $\tauf$ in units of the Lyapunov time for varying reservoir dimension $D_r$ for RAFDA and for LR, for noise contaminated observations with $\eta=0.2$. The error bars denote two standard deviations, estimated from $500$ independent realisations, differing in their randomly drawn internal parameters ($\Win$, $\bin$), training and validation data sets.}
\label{fig.L63_Dr}
\end{figure}

Figure~\ref{fig.L63_Dr} shows the dependency of the forecast time $\tauf$ on the reservoir dimension $D_r$ for fixed training length $N=4,000$, averaged over $500$ realisations, each involving independently drawn random internal parameters $\Win$ and $\bin$ as well as different training and validation data sets. For each reservoir dimension $D_r$ we choose an ensemble with $M=D_r$ members. The forecast performance of both RAFDA and LR increases with the reservoir dimension $D_r$ initially, with a very steep increase for RAFDA between $\log D_r=3$ and $\log D_r=4$, and then saturates for sufficiently large $D_r$. LR saturates for reservoir dimensions $\log D_r \approx 4$ to $\tau_f \approx 1.4$. RAFDA saturates at larger reservoir dimensions. At saturation RAFDA achieves forecast times $\tau_f\approx 3.4$, more than twice than the corresponding values for classical LR. Note that the mean forecast time $\tau_f$ of LR at large reservoir dimensions $D_r\approx 1,500$ ($\log D_r \approx 7.3$) is still smaller with $\tau_f\approx 1.4$ than the mean forecast time of RAFDA for the reservoir dimension $D_r=50$ ($\log D_r\approx  3.9$) with $\tau_f\approx 2.0$. This suggests that for training it is far more important to control the noisiness of the observations than the expressivity of the model. 


Figure~\ref{fig.L63_Dr} also shows that there is a large spread in the obtained forecast times $\tauf$ caused by outliers in both LR and RAFDA; the spread, relative to the mean forecast time $\tau_f$, however, is diminished by approximately $25\%$ on average for RAFDA compared to LR.   

%

\begin{figure}[htbp]
\centering
\includegraphics[width = 0.5\columnwidth]{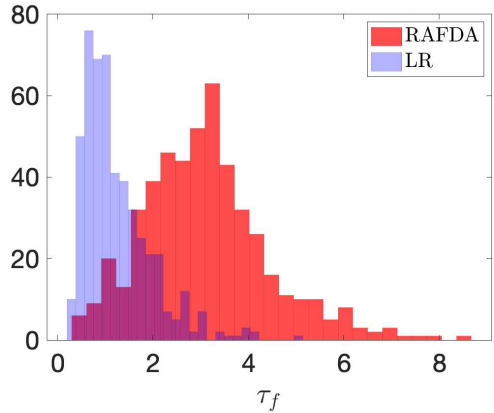}
\caption{Empirical histogram of forecast times $\tauf$ in units of the Lyapunov time for $D_r=1,600$, for noise contaminated observations with $\eta=0.2$.}
\label{fig.L63_Dr_hist}
\end{figure}

To further illustrate the spread of the forecast times, we show in Figure~\ref{fig.L63_Dr_hist} the empirical histogram of the forecast time $\tauf$ for RAFDA and for LR obtained from $500$ independent realisations for a reservoir of size $D_r=1,600$. RAFDA clearly exhibits improved forecast times with a mean of $\tau_f=3.4$ over LR with a mean of $\tau_f=1.4$  Lyapunov time units. Moreover, the large outliers for RAFDA can have forecast times larger than $9$ Lyapunov times, whereas extreme cases for LR only reach up to $5.6$ Lyapunov times. We used here $M=1,600$ ensemble members. The models are trained with $N =4,000$ noisy observations.

In anticipation of the application to higher-dimensional dynamical systems presented in Section~\ref{sec:KS} and \ref{sec:L96} we state that the reservoir dimension $D_r$ for which the forecast skill saturates depends on the dimension $D$ of the underlying system; the higher the dimension of the dynamical system, the higher the expressivity of the model needs to be.

\subsection{Dependency on the training data size $N$}
The size of the training data set needs to be sufficiently large for two distinct reasons.  First of all, the training data set needs to sample sufficiently large  parts of phase space to allow for an application of the learned mapping to unseen data which may evolve into hitherto unexplored areas of phase space. Secondly, the size of the training data set needs to be sufficiently large to allow for good statistical estimation in both the linear regression as well as the data assimilation components (which has its own initial transient period) and hence to ensure a sufficiently accurate estimation of the mappings (\ref{eq:LR}) and (\ref{eq:RAFDA}), respectively. 

\begin{figure}[htbp]
\centering
\includegraphics[width = 0.5\columnwidth]{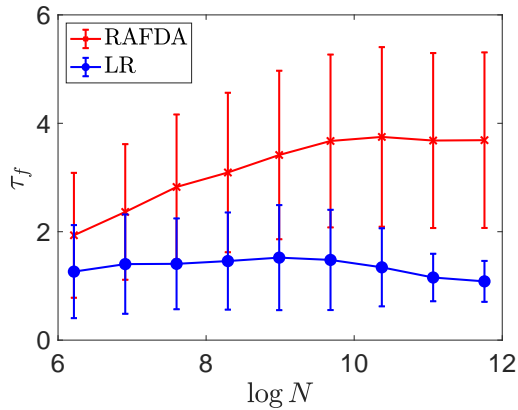}
\caption{Mean of the forecast time $\tauf$ in units of the Lyapunov time for varying lengths $N$ of the training data set for RAFDA and for LR, for noise contaminated observations with $\eta=0.2$. The error bars denote two standard deviations, estimated from $500$ independent realisations, differing in their randomly drawn internal parameters ($\Win$, $\bin$), training and validation data sets.}
\label{fig.L63_N}
\end{figure}

Figure~\ref{fig.L63_N} shows the dependency of the mean of the forecast time $\tauf$ on the size $N$ of a noisy training data set, averaged over $500$ realisations, again  each involving independently drawn random internal parameters $\Win$ and $\bin$ as well as different training and validation data sets. We checked that the analysis fields for the Lorenz-63 system state variables were close to the truth and no ensemble inflation was required. As for the reservoir dimension, the forecast times $\tauf$ increase with increasing length of the training data set $N$ and saturate for sufficiently large training data sets, exhibiting the same twofold improvement of RAFDA over LR as observed in Section~\ref{sec:Dr}. The slight decline in the forecast time for $\log N>10$ for LR is consistent with sampling errors, illustrated by the error bars shown in Figure~\ref{fig.L63_N}. Both LR and RAFDA suffer from outliers in the prediction which gives rise to the variance in forecast times (see also Figure~\ref{fig.L63_Dr_hist}). We observe that the spread does not decrease with increasing training data length $N$ but saturates in conjunction with the mean forecast time $\tauf$. As in the dependency on the reservoir dimension, the spread relative to the mean forecast time is reduced on average by $20\%$ for RAFDA compared to LR.


%
%


\subsection{Dependency on the measurement noise level $\eta$}
\begin{figure}[htbp]
\centering
\includegraphics[width = 0.5\columnwidth]{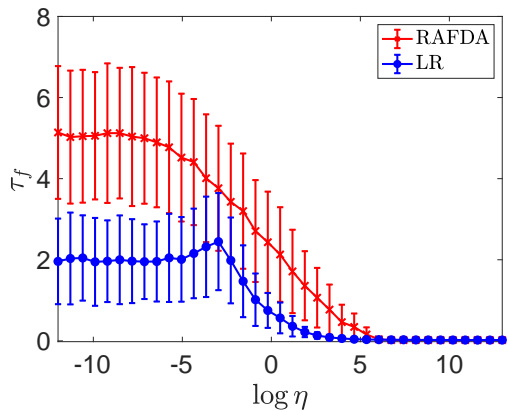}
\caption{Mean of the forecast time $\tauf$ in units of the Lyapunov time for varying noise levels $\eta$ for RAFDA and for LR. The error bars denote two standard deviations, estimated from $500$ independent realisations, differing in their randomly drawn internal parameters ($\Win$, $\bin$), training and validation data sets.}
\label{fig.L63_eta}
\end{figure}

We now test how the forecast capabilities depend on the observational noise level $\eta$. Figure~\ref{fig.L63_eta} shows the dependency of the forecast time $\tauf$ on the noise level for fixed training data length and reservoir dimension, averaged over $500$ realisations, each involving independently drawn random internal parameters $\Win$ and $\bin$ as well as different training and validation data sets. For sufficiently small noise levels $\eta<0.0067$ ($\log \eta<-5$) RAFDA asymptotes to a mean forecast time of $\tauf\approx 5$  whereas LR asymptotes to only $\tauf\approx 2$, albeit for a longer range in noise levels. 
The range of constant forecast times is followed upon increasing the noise level by an exponentially decaying range, before the models lose any forecast skill with $\tauf\to 0$. The exponential behaviour of the forecast time $\tauf$ with respect to the observational noise strength $\eta$ is consistent with the sensitivity of chaotic dynamical systems with respect to their initial condition and the exponential separation of nearby trajectories. To corroborate this interpretation we have confirmed that the same forecast times are achieved when simultaneously doubling the data size to $N=8,000$ and the noise level $\eta$.

The forecast time $\tau_f$ varies non-monotonically with the observational noise strength $\eta$ for LR. We observe an increase in $\tau_f$ for $\log \eta \approx -3$. The value of the observational noise $\eta$ for which this increase in the $\tau_f$ occurs depends on $\beta$ and decreases for decreasing values of $\beta$ (not shown). Recall that LR has been devised under the assumption of model error, with covariance proportional to $\beta$, without consideration of eventual observational error. It seems that a time series with a specified observational noise can be approximated by an equivalent model error, and hence allows for an adequate application of LR for this particular observational noise strength.

To further illustrate that standard random feature maps and LR are not suited to learn the dynamics from noisy observations, we tested that simple denoising does not lead to an improved forecasting skill of LR. By applying a simple denoising algorithm of a moving average whereby each observation $\uobs(t_i)$ is replaced by $\tfrac13(\uobs(t_i-\Delta t)+\uobs(t_i)+\uobs(t_i+\Delta t))$ to average out the noisy signal \cite{KantzSchreiber}, yields a mean forecast time of $\tau_f=1.6$ for $\eta=0.2$, compared to $\tau_f=3.3$ for RAFDA, when averaged over $500$ realisations as above. Similarly, the RMS error at lead time $t=1$ is $9.1$ compared to $3.9$ for RAFDA. It is likely that a more refined denoising scheme could boost the performance of LR; but such a scheme would be problem dependent and would require additional tuning. \\

The reader may wonder why RAFAD outperforms LR for $\eta\to 0$ as for $\eta=0$. If the output data were lying in the span of the random feature maps, then linear regression would be the optimal solution minimising (\ref{eq:RR}) with $\mathcal{L}(\W_{\rm{LR}})=0$. We remark that LR with $\beta>0$ assumes an intrinsic model error (\ref{eq:model_error}) which is absent from our deterministic Lorenz-63 system. We have tested for $\beta=10^{-11}$ and several larger values of $\beta$ that $0<\mathcal{L}(\W_{\rm{LR}})< \mathcal{L}(\W_{\rm{RAFDA}})$ for the training data but nevertheless RAFDA generates surrogate models which consistently exhibit larger forecast times $\tau_f$ when applied to unseen data (not shown). Hence, despite the fact that RAFDA performs suboptimally in minimising the cost function (\ref{eq:RR}) it generalises much better to unseen data. The fact that the EnKF does not yield optimal estimates for nonlinear problems has long been known (see, for example, \cite{LeGlandEtAl11}).


%
%


\subsection{Application to ensemble forecasting}
In chaotic dynamical systems a single forecast is to a certain degree meaningless as small uncertainties in the initial condition can lead to widely differing forecasts at a later time. Probabilistic forecasts provide a more appropriate framework for forecasting. In particular, ensemble forecasting, whereby a Monte Carlo estimate of the underlying probability density function is provided by running an ensemble of initial conditions, are now widely used in numerical weather forecasting issuing both the most probable forecast and its associated uncertainty \citep{Epstein69,Leith74,LeutbecherPalmer08}. The quality of such ensemble forecasts crucially depends on how the ensembles are generated. There exist several strategies using singular vectors \citep{Lorenz65,Palmer93}, bred vectors \citep{TothKalnay93,TothKalnay97,GigginsGottwald19}, analysis ensembles from ensemble Kalman filters \citep{WangBishop03,BuizzaEtAl05}, and more recently analogs \citep{AtenciaZawadzki17}. Here we show that ensembles obtained from RAFDA provide reliable forecast ensembles to be used in probabilistic forecasts. A good probabilistic forecasts is not necessarily one with the smallest root-mean-square (RMS) error (consider a probability density function with disjoint support, then the mean may not be even an actual physical state), but rather one where each ensemble member has equal probability of being closest to the truth. Such ensembles are called reliable. We probe the reliability here with the continuous ranked probability score (\textrm{CRPS}) \citep{hersbach00,Wilks}. The \textrm{CRPS} is defined for a lead time $\tau$ as
\begin{align}
\mathrm{CRPS}(\tau) = \frac{1}{D}\sum_{k=1}^{D}\int_{-\infty}^{\infty} \Big[F(u;\tau,k) - F_{\rm{truth}}(u;\tau,k)\Big]^2 \,\mathrm{d}u ,
\end{align}
where $F$ and $F_{\rm truth}$ are the cumulative probability distributions of the forecast ensemble and truth, respectively, which are estimated as
\begin{align}
F(u;\tau,k) &= \frac{1}{M}\sum_{m=1}^M \Theta\Big(u - u^{(m,k)} (\tau)\Big),\\
F_{\rm truth}(u;\tau,k) &= \Theta\Big(u-u_{\rm{truth}}^{(k)}(\tau)\Big),
\end{align} 
where $\Theta(x)$ is the Heaviside function with $\Theta(x)=0$ for $x<0$ and $\Theta(x)=1$ otherwise, and $u^{(m,k)}(\tau)$ 
denotes the $k$-component of the $m$th ensemble member $\u^{(m)}(\tau)$ at forecast time $\tau>0$. Smaller values of CRPS indicate better reliability.

\begin{figure}[htbp]
\centering
\includegraphics[width = 0.5\columnwidth]{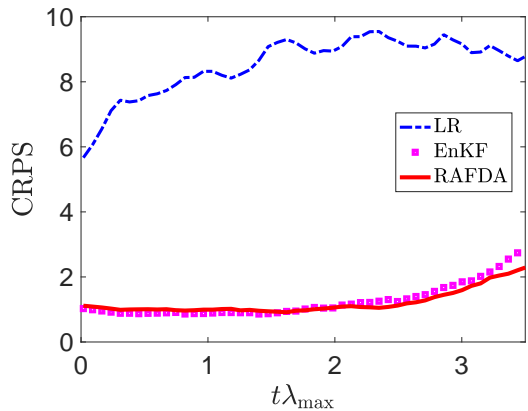}
\caption{$\CRPS$ as a function of the lead time $\tau$ for RAFDA, LR and EnKF ensembles.}
\label{fig.L63_CRPS}
\end{figure}

We consider here ensembles obtained in an EnKF data assimilation where the forecast model is given by the surrogate model obtained from either RAFDA or from LR. Both surrogate models and their associated weight matrices were obtained prior to and independently from the data assimilation cycles using a training set of length $N=250,000$, consisting of noisy observations with $\eta = 0.2$ sampled every $\Delta t=0.02$ model time units. The LR and RAFDA models are kept fixed and no further parameter estimation occurs during the data assimilation. We use an observation time of $\Delta t=0.02$ for the independent data assimilation cycles used to generate the ensembles; the same time interval used for training both LR and RAFDA. We further compare with results obtained from a traditional EnKF ensemble where the full Lorenz-63 model (\ref{eq:L63}) is employed as forecast model. We perform $500$ independent analysis cycles, allowing for each a burn-in period of $2,000$ analysis cycles before the analysis error and the \textrm{CRPS} are evaluated.  RAFDA performs equally well as the full EnKF which uses the full Lorenz-63 model (\ref{eq:L63}) as forecast model, with analysis errors of $2.4$ and $2.3$, respectively. The LR model was not able to produce good analysis with a remarkably larger analysis error of $11.0$. Figure~\ref{fig.L63_CRPS} shows $\CRPS(\tau)$, averaged over the $500$ independent realisations. Whereas LR ensembles cannot be classified as reliable, RAFDA ensembles are shown to be as reliable as EnKF ensembles. 
 
%
%


\section{Partial differential equations: Kuramoto-Sivashinsky equation}
\label{sec:KS}
We now consider artificially generated time series obtained from a numerical simulation of the Kuramoto-Sivashinsky equation \citep{KuramotoTsuzuki75,KuramotoTsuzuki76,Sivashinsky77,SivashinskyMichelson80} 
\begin{align}
u_t + u u_x + u_{xx} + u_{xxxx} = 0
\label{e.KS}
\end{align}
with periodic boundary conditions. For system length $L=22$ the Kuramoto-Sivashinsky equation is chaotic with a maximal Lyapunov exponent of $\lambda_{\rm{max}}=0.043$ \citep{EdsonEtAl19}. We remark that although the Kuramoto-Sivashinsky equation is a partial-differential equation, its dynamics evolves on a finite dimensional manifold \citep{Temam}. 

We generate observations $\uobs_n $ by integrating (\ref{e.KS}) using a pseudo-spectral Crank-Nicolson scheme where the nonlinearity is treated with a second-order Adams-Bashforth scheme. We employ a temporal discretization step of $\delta t =0.001$ and use $64$ spatial grid points, and sample every $\Delta t = 0.25$ time units to obtain observations $\uobs_n \in \R^{64}$. An initial transient period of $25 \times 10^6$ time units is discarded to ensure that the dynamics has settled onto the attractor.

\begin{figure}[htbp]
\centering
\includegraphics[width = 0.49\columnwidth]{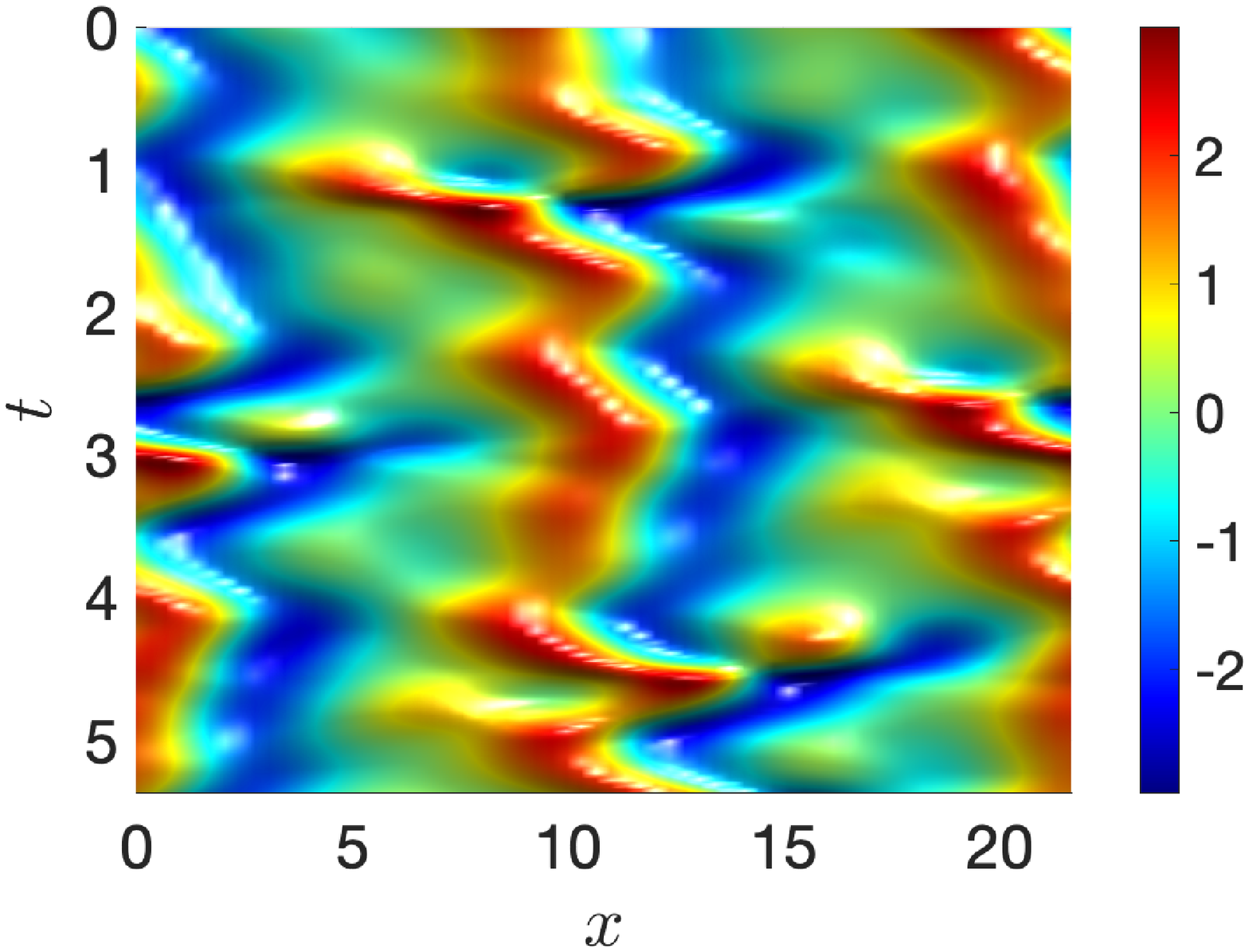}
\includegraphics[width = 0.49\columnwidth]{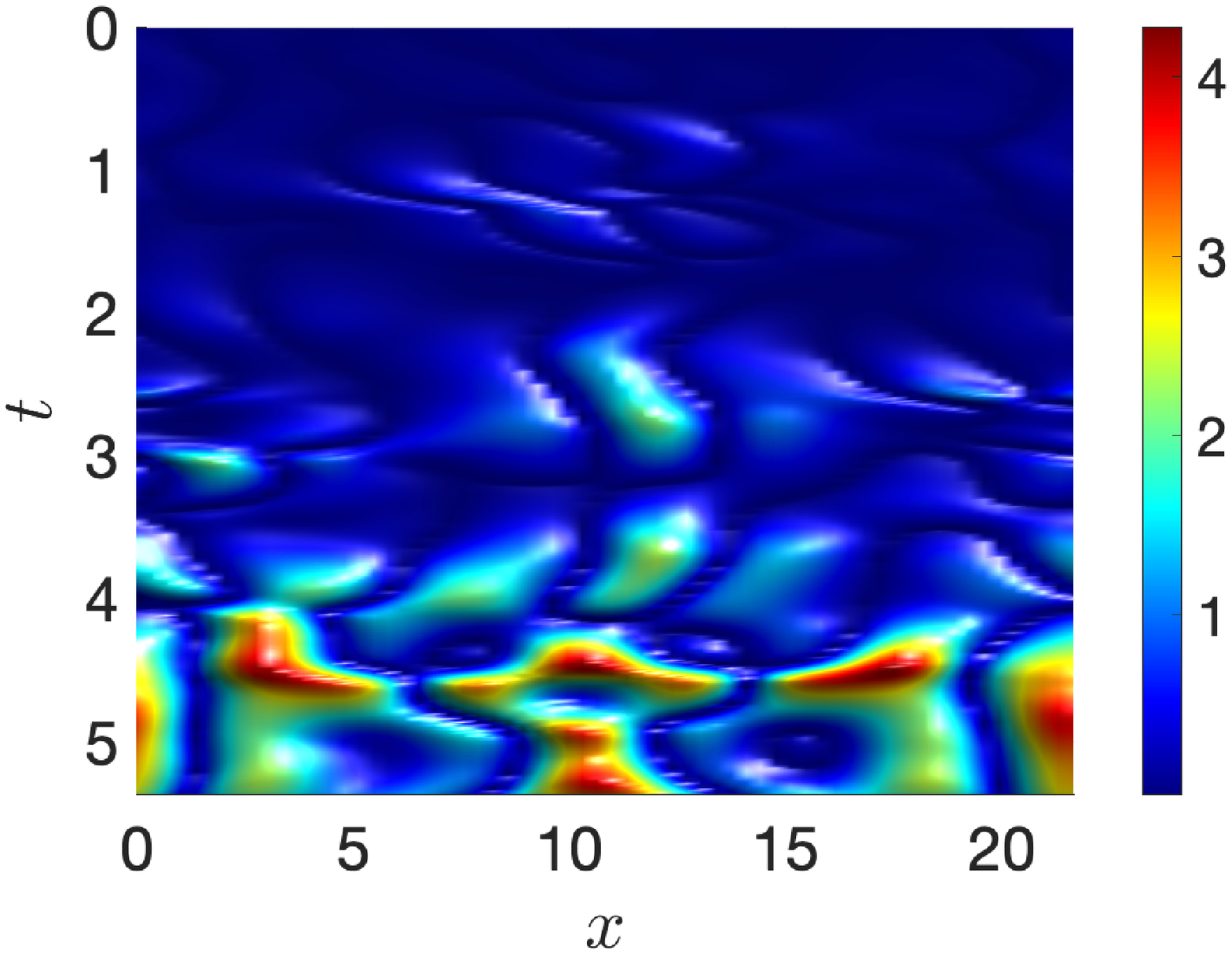}
\caption{Hovm\"oller diagram of the forecast field $u(x,t)$. Left: RAFDA forecast. Right: Root-mean-square error of the RAFDA forecast with respect to the truth. Times are in units of $1/\lambda_{\rm{max}}$.}
\label{fig.KS}
\end{figure}

The internal weight matrix and bias are chosen randomly according to (\ref{eq:wb}) with $w=0.1$ and $b=1$ with a reservoir of size $D_r=2,000$. We consider here only a single realisation of the internal parameters $(\Win,\bin)$. We train RAFDA on a noisy training data set of length $N=70,000$ with $\eta=0.01$. The data assimilation component of RAFDA was executed with $M=1,000$ ensemble members and without any inflation with an initial ensemble chosen according to (\ref{eq:w0}) with $\gamma=6.4 \cdot 10^{-6}$. Figure~\ref{fig.KS} depicts results of the RAFDA forecast showing that reasonable forecasts can be made up to four Lyapunov times. Classical random feature maps with linear ridge regression trained on the same data set only yield forecast horizons of one Lyapunov time (not shown). If LR is trained on noiseless data, it achieves comparable forecast skill to RAFDA (not shown), again illustrating that LR has difficulties with training on noisy data. 

To explore the capabilities of RAFDA to deal with higher-dimensional systems more careful numerical simulations would be needed. We have refrained here from optimizing the reservoir weights $w$ and $b$ and the reservoir dimension $D_r$ to achieve larger forecast horizons, and have only shown a single realization which shows that higher-dimensional problems can be addressed within our framework.

%


\section{Closure models: Multi-scale Lorenz-96 system}
\label{sec:L96}
We consider now the problem of model closure in multi-scale systems. In multi-scale systems one is typically only interested in the dynamics of the large-scale slow variables. Moreover, one typically only has access to observations of the resolved slow dynamics. The objective of model closure, also known as the subgrid-scale parametrization problem, is to find a self-consistent dynamical model for the resolved slow large-scale variables. The effective model for the slow large-scale dynamics allows for a larger time step in the numerical simulation compared to the prohibitively small time steps required in the full stiff multi-scale system. 

We consider here the multi-scale Lorenz-96 system for $K$ slow variables $X^{(k)}$ which are each coupled to $J$ fast variables $Y^{(j,k)}$, given by 
\begin{align}
\frac{d}{dt}X^{(k)} &= -X^{(k-1)}(X^{(k-2)}-X^{(k+1)})-X^{(k)}+F-\frac{hc}{d}\sum\limits_{j=1}^{J}Y^{(j,k)}, 
\label{e.L96_X}\\
\frac{d}{dt}Y^{(j,k)} &= -cbY^{(j+1,k)}(Y^{(j+2,k)}-Y^{(j-1,k)})-cY^{(j,k)}+\frac{hc}{d}X^{(k)},  
\label{e.L96_Y}
\end{align}
with cyclic boundary conditions $X^{(k+K)}=X^{(k)}$, $Y^{(j,k+K)}=Y^{(j,k)}$ and $Y^{(j+J,k)} = Y^{(j,k+1)}$, giving a total of $D=K(J+1)$ variables. This set of equations was introduced as a caricature for the midlatitude atmosphere dynamics \citep{Lorenz96}. The variables $X^{(k)}$ model large scale atmospheric fields arranged on a latitudinal circle, such as synoptic weather systems. Each of the $X^{(k)}$ variables is connected to $J$ small-scale variables $Y^{(j,k)}$. The time scale separation is parametrized by the coefficient $c$, and the ratio of the amplitudes of the large-scale and the small-scale variables is controlled by $b$. The coupling between the slow and fast variables is controlled by the parameter $h$. Both large-scale and small-scale variables evolve, when uncoupled, according to nonlinear transport and linear damping; the large-scales are additionally subjected to external forcing $F$. We choose here as parameters $K=8$ and $J=32$, leading to a total of $256$ variables, and $F=20$, $c=d=10$ and $h=1$ as in \cite{Wilks05,ArnoldEtAl13}. The choice $c=d=10$ implies that the fast variables  $Y^{(j,k)}$ fluctuate with a $10$ times higher frequency and with an approximately $10$ times smaller amplitude when compared to the slow variables $X^{(k)}$. Setting the coupling constant $h=1$, corresponds to strong coupling where the dynamics is driven by the fast sub-system \citep{HerreraEtAl10}.

Our aim is to provide a reduced model of the form
\begin{align}
\frac{d}{dt}X^{(k)} &=G^{(k)}(\X) + \psi(X^{(k)}),
\label{e.L96_closure}
\end{align}
with $\X = (X^{(1)},\ldots,X^{(K)})^{\rm T}$, $G^{(k)}(\X)= -X^{(k-1)}(X^{(k-2)}-X^{(k+1)})-X^{(k)}+F$, and where the closure term $\psi(x)$ parametrizes the effect of the fast unresolved dynamics. Here the closure term $\psi(x)$ will be learned by RAFDA. Note that one and the same closure map can be applied at all grid points due to the spatial homogeneity  of the problem. The application we have in mind is that a scientist has available a physics-based model for the resolved slow large-scale dynamics, i.e. the general form of the equation (\ref{e.L96_closure}) and the expression for the resolved vectorfield $G^{(k)}(\X)$, but does not have access to the unresolved fast small-scale model, and the closure model needs to be inferred from noisy observations. 

In the case when the full multi-scale model (\ref{e.L96_X})--(\ref{e.L96_Y}) is available and noiseless data of both the slow and fast variables are available, the closure problem has been attempted outside the scope of machine learning in \cite{Wilks05,ArnoldEtAl13}. Therein deterministic parametrizations were proposed by data-fitting of the closure term to the resolved variables with polynomials to obtain  
\begin{align}
\psi_{\rm{Wilks}}(x) &= 0.262+1.45 x- 0.0121 x^2 - 0.00713 x^3 + 0.000296 x^4
\label{eq:Wilks}
\\
\psi_{\rm{AMP}}(x) &= 0.341+1.3 x-0.0136 x^2-0.00235 x^3.
\label{eq:Arnold}
\end{align}
Here we assume we are only given noisy observations of the slow variables $\mathbf{X}_n$ at discrete times $t_n = n\Delta t$. In LR, we set $\u_n = \X_n$, $n=0,\ldots,N$, and the closure mapping $\psi(x)$ is learned directly from the data via finite-differencing. More specifically, the closure is performed by minimising the difference between the random feature map approximation of the closure term $\psi (x) = \phi(x)$ over all grid points $x = X^{(k)}$, and the approximation of the closure term provided by the observations 
\begin{align*}
\frac{X_n^{(k)}-X_{n-1}^{(k)}}{\Delta t} - G^{(k)}(\X_{n-1}). 
\end{align*}
This finite differencing, however, constitutes a bad estimator of the closure term $\psi(x)$ for noisy observations of the slow variables. In the following we therefore only present results for LR in the case of zero measurement error (i.e. the same setting in which the polynomial fits (\ref{eq:Wilks}) and (\ref{eq:Arnold}) were obtained). RAFDA, on the other hand, employs as the forecast model within the data assimilation training phase the reduced model (\ref{e.L96_closure}) where $\psi(x)$ is estimated using random feature maps $\psi(x) =\W \tanh(\Win x + \b_{\rm in})$ (cf. (\ref{eq:r}) and (\ref{eq:Wr})) where $\W\in \R^{1\times D_r}$, $\Win\in \R^{D_r\times 1}$, and $\b_{\rm in} \in \mathbb{R}^{D_r}$. RAFDA is trained on $N =100,000$ noisy observations of all components $X^{(k)}$ of $\X$ with variance $\eta=0.02$ sampled at $\Delta t=0.0005$, and an initial ensemble chosen according to (\ref{eq:w0}) with $\gamma=0.01$, centred around the prior provided by LR. We employ inflation $\alpha=1+0.1 \Delta t$. We chose a reservoir of size $D_r=1,000$ and draw the internal parameters using $w=0.05$ and $b=1$. The validation is performed by propagating the reduced model (\ref{e.L96_closure}) with the learned mapping $\psi(x)$ added at each grid point in an Euler time-stepping method with step-size $\delta t=0.005$ for a total of $10^6$ time steps. 

\begin{figure}[htbp]
\centering
\includegraphics[width = 0.5\columnwidth]{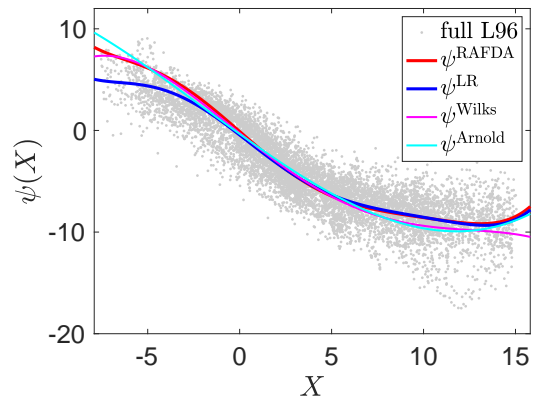}
\caption{Closure terms $\psi=\psi(x)$ for the multi-scale L96-system (\ref{e.L96_X})-(\ref{e.L96_Y}) estimated using various parametrizations.}
\label{fig.L96_psi}
\end{figure}

\begin{figure}[htbp]
\centering
\includegraphics[width = 0.5\columnwidth]{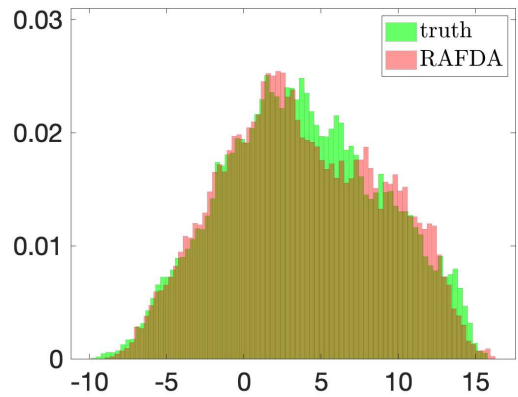}
\caption{Empirical probability density function of the slow variables $X_k$ of the full multi-scale L96-system (\ref{e.L96_X})-(\ref{e.L96_Y}) and of the RAFDA closure model.}
\label{fig.L96_pdf}
\end{figure}

Figure~\ref{fig.L96_psi} shows the closure term $\psi(x)$ estimated from RAFDA, which was trained with noisy data with $\eta=0.02$,  and from LR and from the polynomial fits (\ref{eq:Wilks})--(\ref{eq:Arnold}), which were all obtained from noise free data. Table~\ref{table:L96} lists the associated relative errors of the mean and the variance of $\X$ for the various parametrization schemes used. The closure model provided by RAFDA outperforms both the closure of LR and the ones given by (\ref{eq:Wilks})--(\ref{eq:Arnold}). We show in Figure~\ref{fig.L96_pdf} the empirical probability density function of a validation time series, created independently of the training data set. It is seen that the reference empirical probability density function of the full multi-scale L96 system assigns more probability to higher values of $\X$. We conjecture that this underdispersiveness of the closure models stems from their assumed deterministic nature. The deterministic closures schemes considered here ignore the diffusive effect of the fast variables which arises for finite time-scale and spatial-scale separation as fluctuations around the deterministic mean behaviour. As a first step towards a full stochastic parametrization one could use RAFDA to learn the propagator of the mean $E^{(k)}=\sum\nolimits_{j=1}^{J}Y^{(j,k)}$ alongside the mapping $\psi (x)$. 



%
\begin{table}[htbp]
\caption{Relative error of the mean $\mu$ and the variance $\sigma^2$ of the slow variables $\X$ for the various parametrization schemes.}
\label{table:L96}
\begin{center}
\begin{tabular}{ccc}
  &  $\mu$ & $\sigma^2$\\
\hline
RAFDA & $0.0056$ & $0.019$\\
LR &  $0.029$ & $0.028$\\
Wilks &   $0.033$ & $0.065$\\
Arnold et al. &   $0.016$ & $0.063$\\
\hline
\end{tabular}
\end{center}
\end{table}
%


\section{Discussion and outlook}
\label{sec:discussion}
We have proposed a new data-driven physics-agnostic forecast model, coined RAFDA, which combines random feature maps with DA. The method is designed to provide cheap surrogate mappings to be used in forecasting or to construct closure models when no {\em{a priori}} knowledge of the evolution equations is available and the underlying system is only accessible via noisy observations. The linear coefficients of the random feature map are sequentially updated incorporating incoming observations using an EnKF, which itself uses the random feature map as its forecast model. We have tested the forecast capabilities in several situations for ordinary and partial differential equations. For the Lorenz-$63$ model and the Kuramoto-Sivashinsky equation we were able to achieve a forecast horizon of several Lyapunov times. We investigated the influence on the data length, the number of feature maps and the noise strength on the forecast horizon. Depending on the dimension of the underlying dynamical system a larger number of  random feature maps and larger training data sets are required. Once trained, however, the computational cost to run the surrogate model can be significantly cheaper than numerical integration of the original full model. For example, for the Kuramoto-Sivashinsky model we used a discretization time step of $\delta t=10^{-3}$ whereas the surrogate RAFDA model was trained on a time series with $\Delta t = 0.25$, which implies a computational gain in running time of $\Delta t/\delta t=250$. However, we stress that the real advantage of RAFDA becomes apparent when the model is unknown. 

Besides providing remarkable forecast skills for individual trajectories, we showed that ensembles generated in data assimilation cycles which use a pre-trained RAFDA forecast model are reliable which makes them attractive candidates for probabilistic forecasting where one would like to issue an uncertainty quantification of a forecast. Furthermore, we showed that in the multi-scale Lorenz-$96$ model RAFDA allowed us to find a Markovian closure model for the resolved variables having only observed its slow variables. 


We remark that our methodology of sequentially updating the parameters with the incoming observations using data assimilation is not restricted to EnKFs, but one may instead employ other linear or nonlinear filtering methods such as \cite{CM19}; equally to estimate the parameters $\W$ one is not restricted to linear regression but may employ nonlinear regression methods if desired.

An open question is how to choose the random parameters of the feature map. The universal approximation property ensures that any continuous function can be well approximated by a linear combination of the features, but it does not provide any statement on how to choose the hyper-parameters and on how many features are needed to achieve good appproximation. We have shown that the forecasts capabilities sensitively depend on the random hyper-parameters and that parameters which lead to good approximation in the noiseless case may fail to approximate the forecast map when trained on noisy data. A natural extension is to determine the optimal parameters $\Win$ and $\bin$ simultaneously with $\W$, possibly using again an ensemble Kalman filter. Rather than random features which are related to RKHS, the machine-learning framework would then consist of a two-layer network \citep{ChizatBach18,MeiEtAl18,RotskoffVandenEijnden18,SirignanoSpiliopoulos20,EEtAl19}. This extension is planned for further research.

Further, recent developments on operator-valued kernels in the context of random feature maps has shown that the number of learnable parameters can be reduced to $D_r$, independent of the dimension $D$ of the underlying system \cite{KadriEtAl16,NelsenStuart20}. 
The methodology developed in this paper easily generalizes to such vector- or operator-valued random feature maps. However, it remains to be investigated whether or not the required number $D_r$ of feature maps remains indeed independent of the dimension $D$ of the
underlying dynamical systems unless the random feature maps are specifically designed for the problem at hand. This poses an interesting
question for further research, in particular in view of dealing with high-dimensional systems. The same applies to the application of
localization to the EnKF. Provided more information is available on the correlation structure of the unknown parameters $\W$, a more refined localization procedure could be employed. 

The numerical experiments in this paper have  all been conducted under an ideal twin setting where the underlying dynamics is
deterministic and no random or systematic model errors had to be considered. The robustness of the proposed RAFDA with regard
to such random or systematic perturbations will be the subject of future work. The same applies to a more systematic and theoretical investigation of the impact of the various algorithmic parameters on the behavior of RAFDA within the ideal twin setting and beyond.


\section*{Acknowledgments}
The work of SR has been partially funded by
Deutsche Forschungsgemeinschaft (DFG, German Science Foundation) - SFB 1294/1 - 318763901.


\bibliographystyle{abbrvnat}


\end{document}